\documentclass[12pt]{article}
\usepackage{a4,amsmath,amssymb,cite,graphicx,multirow,slashbox,cancel}
\def\Bbb{\mathbb}
\def\BZ{\Bbb Z} 
\def\BC{\Bbb C} 
   \def\BH{\mathbb{H}}

\def\Tr{\mathrm{Tr}}

\newcommand{\Sp}{\textrm{Sp}}
\newcommand{\etabox}[2]{\underset{\ ~#2}{#1\ \framebox[15pt]{\phantom{a}}}}
\catcode`@=11 \@addtoreset{equation}{section} \catcode`@=12

\begin{document}
\bibliographystyle{utphys}
\begin{titlepage}
\renewcommand{\thefootnote}{\fnsymbol{footnote}}
\noindent
{\tt IITM/PH/TH/2010/5}\hfill
{\tt arXiv:1006.3472 [hep-th]} \\[4pt]
\mbox{}\hfill 
\hfill{\fbox{\textbf{v2.0; April 27, 2011 }}}

\begin{center}
\large{\sf  BKM Lie superalgebras from counting twisted CHL dyons}
\end{center} 
\bigskip 
\begin{center}
Suresh Govindarajan\footnote{\texttt{suresh@physics.iitm.ac.in}} \\
\textit{Department of Physics, Indian Institute of Technology Madras,\\ Chennai 600036, INDIA.}
\end{center}
\bigskip
\begin{center}
\textsf{\small Dedicated to the memories of Jaydeep Majumder and Alok Kumar}
\end{center}
\bigskip
\begin{abstract}

Following Sen, we study the counting of (`twisted') BPS states that contribute to  twisted helicity trace indices in four-dimensional CHL models with $\mathcal{N}=4$ supersymmetry. The generating functions of half-BPS states, twisted as well as untwisted, are given in terms of multiplicative eta products with the Mathieu group, $M_{24}$, playing an important  role. These multiplicative eta products enable us to construct Siegel modular forms that count twisted quarter-BPS states. 

The square-roots of these Siegel modular forms turn out be precisely a special class of Siegel modular forms,  \textit{the dd-modular forms}, that have been classified by Clery and Gritsenko. We show that each one of these dd-modular forms arise as the Weyl-Kac-Borcherds denominator formula of a rank-three Borcherds-Kac-Moody Lie superalgebra. The walls of the Weyl chamber are in one-to-one correspondence with the walls of marginal stability in the corresponding CHL model for twisted dyons as well as untwisted ones. This leads to \textit{a periodic table} of  BKM Lie superalgebras  with properties that are consistent with physical expectations. 
\end{abstract}
\begin{center}
\texttt{appeared as JHEP 05 (2011) 089}
\end{center}
\end{titlepage}
\setcounter{footnote}{0}
\section{Introduction}

The microscopic counting of black hole degeneracy in four-dimensional string theories with $\mathcal{N}=4$ supersymmetry has seen enormous progress in the past few years. The prototypical example is furnished by type II string theory compactified on $K3\times T^2$ or equivalently, the heterotic string compactified on $T^6$. In this case,  the  degeneracy of electrically charged $\tfrac12$-BPS states is generated by the  $SL(2,\mathbb{Z})$ modular form of weight 12, $\Delta(\tau)=\eta(\tau)^{24}$, while the degeneracy of (dyonic) $\tfrac14$-BPS states is the  $Sp(2,\mathbb{Z})$ Igusa cusp form of weight $10$, $\Phi_{10}(\mathbf{Z})$\cite{Dijkgraaf:1996it}. There are two natural extensions of this construction.
\begin{enumerate}
\item Carry out  the analogous counting to other models with $\mathcal{N}=4$ supersymmetry such as the CHL orbifolds\cite{Sen:2005ch,Jatkar:2005bh,David:2006ji,David:2006yn,David:2006ud,Dabholkar:2008zy,Govindarajan:2009qt}
\item  Carry out the counting for a subset of states(that we call \textit{twisted} BPS states) that contribute to a twisted index first considered by Sen\cite{Sen:2009md}.
\end{enumerate}
One can, of course, combine both the above extensions and count twisted BPS states in the CHL orbifolds\cite{Sen:2010ts}.  The first question that this paper addresses is the following:
Are there modular forms that are the  analogs of the modular forms $\Delta(\tau)=\eta(\tau)^{24}$ and $\Phi_{10}(\mathbf{Z})$ for $\mathbb{Z}_M$-twisted BPS states in CHL $\mathbb{Z}_N$-orbifolds. An observation in \cite{Govindarajan:2009qt} indicates that the  Mathieu group, $M_{24}$, should play an important role in this construction. 

The square-root of the Igusa cusp form, $\Delta_5(\mathbf{Z})$, appears as the Weyl-Kac-Borcherds denominator identity for a particular rank-three Borcherds-Kac-Moody (BKM) Lie superalgebra\cite{Nikulin:1995}. It has been shown that walls of the Weyl chamber of this BKM Lie superalgebra gets mapped to the walls of marginal stability\cite{Sen:2007vb} across which the degeneracy of $\tfrac14$-BPS states jumps due to the non-existence of two-centered black holes on one side of a wall\cite{Cheng:2008fc}. A similar relation has been observed between the square-roots of the genus-two modular forms that count $\tfrac14$-BPS states and other rank-three BKM Lie superalgebras for CHL $\mathbb{Z}_N$-orbifolds for $N=2,3,4$\cite{Cheng:2008kt,Govindarajan:2009qt}. Further, it has been observed that the BKM Lie superalgebras constructed in \cite{Govindarajan:2008vi} play an identical role for $\mathbb{Z}_M$-twisted $\tfrac14$-BPS states in type IIA string theory on $K3$. Is there a Lie algebraic structure associated with  twisted dyons in the  CHL $\mathbb{Z}_N$-orbifolds? This is the second question that is addressed in this paper.

\subsubsection*{Summary of results}

We summarize the main results of this paper.
\begin{enumerate}
\item We provide a derivation of the map, Eq. \eqref{cycleshapemap}, that relates cycle shapes ($M_{24}$ conjugacy classes) to  multiplicative eta-products. These eta-products are the generating functions of electrically charged twisted and untwisted $\tfrac12$-BPS states (see appendix \ref{derivation}).
\item We construct Siegel modular forms, $\Phi^{(N,M)}_k(\mathbf{Z})$, of the paramodular group, $\Gamma_t(P)$ ($t=(N.M)$ and $P=\textrm{max}(N/t, M/t)$)
using the additive lift of a Jacobi form. The modular forms are generating functions of $\BZ_M$-twisted $\tfrac14$-BPS states in the CHL $\BZ_N$-orbifold. 
\item The square-root of the modular forms are the dd-modular forms of Clery and Gritsenko. The sum and product representations  of \textit{all} dd-forms are the Weyl-Kac-Borcherds (WKB) denominator formulae for a family of rank-three Lorentzian Kac-Moody Lie superalgebras, $\mathcal{G}_N(M)$. These BKM Lie superalgebras are summarised in Table \ref{periodictable}.
\end{enumerate}
\begin{figure}[hbt]
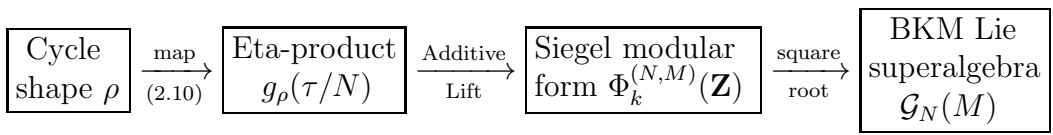

$$
\fbox{\parbox{1.4cm}{~Cycle \\ \mbox{}shape $\rho$}}
\xrightarrow[ \eqref{cycleshapemap}]{\textrm{map}} \fbox{\parbox{2.2cm}{Eta-product \\ \hspace*{0.4cm}$g_\rho(\tau/N)$}}
\xrightarrow[\textrm{Lift}]{\textrm{Additive} }\fbox{\parbox{2.9cm}{Siegel modular\\ form $\Phi^{(N,M)}_k(\mathbf{Z})$}}
\xrightarrow[\textrm{root}]{\textrm{square}}\fbox{\parbox[c]{2.3cm}{~~BKM Lie \\
superalgebra $\mbox{~~~}\mathcal{G}_N(M)$}}
$$
\caption{A pictorial summary of results}\label{pictorialsummary}
\end{figure}

\subsubsection*{Organization of the paper}

After the introductory section, in section 2, we summarize known results  about modular forms that count untwisted BPS states in CHL models. In section 3, we extend these considerations to construct modular forms that  generate the degeneracies of twisted BPS states in the CHL models for twists that are  symplectic automorphisms of K3. These modular forms are shown to have properties that are compatible with macroscopic considerations, wall-crossing  as well as S-duality. In section 4, we show that the square-root of each of these modular forms are the WKB denominator identity of a BKM Lie superalgebra. These BKM Lie superalgebras are shown to have properties that are compatible with physical expectations. We end in section 5 with some concluding remarks. The appendices contain important technical details. Appendix A provides a derivation of the map that relates cycle shapes to eta-products. Appendix B gives the details about the paramodular group, its modular forms and their characters. We also prove the invariance of the constructed modular forms under extended S-duality. Appendix C provides product formulae for the modular forms in fairly explicit form. Appendix D gives the (partial) Fourier expansion of two modular forms $Q_1$ and $\widetilde{Q}_1$ where we explicitly track the simple real roots of two BKM Lie superalgebras.

\section{Counting BPS states in the CHL models}

The CHL models that we consider will be $\mathbb{Z}_N$ orbifolds of type IIA string theory compactified on $K3\times S^1\times \widehat{S}^1$ (or equivalently asymmetric orbifolds of the heterotic string compactified on $T^4\times S^1\times  \widehat{S}^1$)\cite{Chaudhuri:1995dj}. The $\mathbb{Z}_N$ group is generated by the simultaneous symplectic automorphism, $g$, of K3 (of order $N$) and a shift of order $N$, $\tilde{g}$, on the $S^1$-circle. The vector multiplet moduli space is given by
\begin{equation}
\left( \Gamma_1(N)\Big\backslash\frac{SL(2)}{U(1)}\right) \times \left(SO(6,p;\BZ)\Big\backslash\frac{SO(6,p)}{SO(6)\times SO(p)}\right)\ ,
\end{equation}
where $p$ is determined from the orbifold action; $SO(6,p;\BZ)$ is the T-duality symmetry group and $\Gamma_1(N)\subset PSL(2,\BZ)$ 
is the S-duality symmetry group which acts on the heterotic axion-dilaton, $\lambda$, as
\begin{equation}
\lambda \longmapsto \frac{a\lambda+b}{c\lambda+d} \ , \quad \begin{pmatrix} a & b \\ c & d \end{pmatrix} \in \Gamma_1(N)\ .
\end{equation}
There is also a parity transformation that enlarges the modular group from $PSL(2, \BZ)$ to $PGL(2, \BZ)$ and acts on the charges and the  heterotic axion-dilaton\cite{Cheng:2008fc}  
\begin{equation}
w: \left( \begin{array}{c}\mathbf{q}_e \\ \mathbf{q}_m \end{array} \right) \rightarrow  \left( \begin{array}{c}\mathbf{q}_e \\ -\mathbf{q}_m \end{array} \right), \quad \lambda \rightarrow \bar{\lambda}\ .
\end{equation}
On adding the parity symmetry to  the S-duality group, $\Gamma_1(N)$, one obtains   the `extended S-duality symmetry group', $\widehat{\Gamma}_1(N)$\cite{Cheng:2008fc,Cheng:2008kt}. For $N\leq 4$, the group $\widehat{\Gamma}_1(N)$ is generated by the following three generators\cite{Cheng:2008kt,Govindarajan:2009qt}:
\begin{equation}\label{Sdualitygenerators}
\gamma^{(N)} = \begin{pmatrix} 1& -1 \\ N & 1-N \end{pmatrix} \ , \
\delta=\begin{pmatrix} -1& 1 \\ 0 & 1 \end{pmatrix} \ \textrm{ and } \
w=\begin{pmatrix} 1& 0 \\ 0 & -1 \end{pmatrix} \ .
\end{equation}
The first two matrices generate a dihedral group, Dih$(\mathcal{P}_N)$,  which is of infinite order when $N=4$. This dihedral group turns out to be the symmetry of a polygon (Weyl chamber), $\mathcal{P}_N$. 

The electric and magnetic charges, $(\mathbf{q}_e,\mathbf{q}_m)$, transform as vectors under the $T$-duality group.  The quantization of the charges  in terms of $T$-duality invariants is such that
\begin{equation}
N \tfrac{\mathbf{q}_e^2}2 \in \mathbb{Z}\quad,\quad\mathbf{q}_e\cdot \mathbf{q}_m \in \mathbb{Z}\quad,\quad  \tfrac{\mathbf{q}_m^2}2 \in  \mathbb{Z}\ .
\end{equation}
We will indicate these integers, respectively, by $(n,\ell,m)$ in this paper. It is useful to form the matrix, $\mathcal{Q}$,
\begin{equation}
\mathcal{Q} \equiv \begin{pmatrix}  \mathbf{q}_e^2 & 
\mathbf{q}_e\cdot \mathbf{q}_m \\ \mathbf{q}_e\cdot \mathbf{q}_m&
 \mathbf{q}_m^2 \end{pmatrix}
=\begin{pmatrix} \tfrac{2n}N & \ell \\ \ell & 2 m \end{pmatrix}\quad\ ,
\end{equation}
in terms of which the $S$-duality action is given by 
\begin{equation}
\mathcal{Q} \mapsto \gamma \cdot \mathcal{Q} \cdot \gamma^T \quad \textrm{for } \gamma \in \widehat{\Gamma}_1(N) \ .
\end{equation}
For the $\mathbb{Z}_N$-orbifolds of interest,  one has $ \tfrac12\mathbf{q}_e^2 \geq -\tfrac1N$ and $ \tfrac12\mathbf{q}_m^2 \geq -1$.

\begin{figure}[hbt]
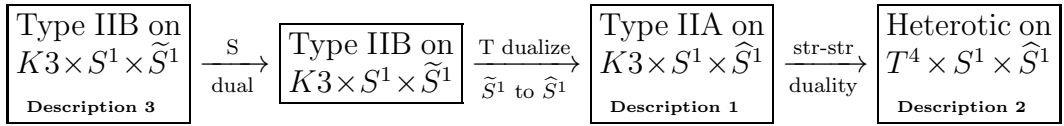

$$
\fbox{\parbox{2.2cm}{Type IIB on \\  $K3\times S^1\times \widetilde{S}^1$\\ \textbf{\tiny \phantom{a}Description~3}}}
\xrightarrow[\textrm{dual}]{~~\textrm{S}~~} \fbox{\parbox{2.2cm}{Type IIB on \\  $K3\times S^1\times \widetilde{S}^1$}}
\xrightarrow[\widetilde{S}^1\textrm{ to } \widehat{S}^1]{\textrm{T dualize} }\fbox{\parbox{2.2cm}{Type IIA on \\  $K3\times  S^1\times \widehat{S}^1$ \\ \textbf{\tiny \phantom{a}Description~1}}}
\xrightarrow[\textrm{duality}]{\textrm{str-str}}\fbox{\parbox[c]{2.2cm}{Heterotic on \\  $T^4\times S^1\times \widehat{S}^1$ \\ \textbf{\tiny \phantom{a}Description~2}}}
$$
\caption{The chain of dualities. The above chain is expected to hold after $\mathbb{Z}_N$-orbifolding of $K3\times S^1$ that leads to the CHL models.}
\label{CHLdualitychain}
\end{figure}

The sequence of dualities which relates the various descriptions of the CHL model are given in the Figure \ref{CHLdualitychain} (taken from \cite{Gopalathesis}). The quantization of charges is specified in the heterotic/CHL string (Description 2) with electric charges being defined as those carried by states of the heterotic string. The microscopic counting of David and Sen is carried out in Description 3 where one has a configuration of D1-D5 branes\cite{David:2006yn} -- this provides a direct derivation of all the modular forms that appear in this paper.

\subsection{Generating functions for degeneracies}

Let $d(n)$ denote the microscopic degeneracy of electrically charged  $\tfrac12$-BPS states. Consider the generating function, $g_\rho(\tau)$, formally defined as follows
\begin{equation}
\frac{16}{g_\rho(\tau/N)} = \sum_{n=-1}^\infty d(n)\ q^{n/N}\ ,
\end{equation}
where $q=\exp(2\pi i \tau)$ and the label $\rho$ will denote a cycle shape as will be discussed later. The $16$ reflects the degeneracy of a single $\tfrac12$-BPS multiplet -- thus $[d(n)/16]$ counts the number of $\tfrac12$-BPS multiplets with electric charge, $\mathbf{q}_e^2/2 = n/N$.

Similarly, let $D(n,\ell,m)$ denote the microscopic degeneracy of dyonic $\tfrac14$-BPS states and define its generating function as follows:
\begin{equation}
\frac{64}{\Phi(\mathbf{Z})} = \sum_{(n,\ell,m)} D(n,\ell,m)\ q^{n/N} r^\ell s^m\ ,
\end{equation}
where $\mathbf{Z} = \left(\begin{smallmatrix} \tau & z \\ z & \sigma \end{smallmatrix}\right)$, $r=\exp(2\pi i z)$ and $s=\exp(2\pi i \sigma)$. The $64$ reflects the degeneracy of a single $\tfrac14$-BPS multiplet -- thus $[D(n,\ell,m)/64]$ counts the number of $\tfrac14$-BPS multiplets with charges $(n,\ell,m)$.

An important point is that both the functions, $g(\tau)$ as well as $\Phi(\mathbf{Z})$ turn out to be modular forms of suitable subgroups of $SL(2,\BZ)$ and $Sp(2,\BZ)$ with $\tau$ ($\mathbf{Z}$) becoming coordinates on the upper-half plane (space).
This enables one to compute the degeneracies, in the limit of large charges, using the saddle point approximation\cite{Dijkgraaf:1996it,Sen:2005ch,Jatkar:2005bh}. These results agree with the macroscopic computation (of the degeneracy) using the Bekenstein-Hawking entropy for $\tfrac14$-BPS extremal black holes\cite{Dijkgraaf:1996it,Jatkar:2005bh} and the Bekenstein-Hawking-Wald entropy for $\tfrac12$-BPS extremal black holes\cite{Dabholkar:2004yr,Dabholkar:2005by,Dabholkar:2005dt,Sen:2005ch}.

\subsection{Eta-products from  $\tfrac12$-BPS degeneracies}

It was shown in ref. \cite{Govindarajan:2009qt} that the $\tfrac12$-BPS degeneracy in the CHL $\mathbb{Z}_N$ orbifolds are given by multiplicative eta-products associated with specific cycle shapes. The cycles shapes correspond to  (conjugacy classes of) elements of the  24-dimensional permutation representation of the Mathieu group $M_{24}$ that also reduce to elements of $M_{23}$\cite{Dummit:1985,Mason:1985}.  The relevant cycle shapes can be obtained from the list given in \cite{Dummit:1985} after imposing the constraints on cycle shapes discussed by Mukai\cite{Mukai:1988} and are reproduced in Table \ref{cycleshapes}. The eta-product, $g_\rho(\tau)$, associated with the  cycle shape, $\rho$ is given by the map
\begin{equation}\label{cycleshapemap}
\rho=1^{a_1}2^{a_2}\cdots N^{a_N} \longmapsto g_\rho(\tau)\equiv \prod_{j=1}^N\eta(j\tau)^{a_j}\ . 
\end{equation}
The weight of the eta-product is given  by $\tfrac12 \sum_s \ a_s\equiv (k+2)$.
A derivation of the eta-products starting from the explicit orbifold action is provided in appendix \ref{derivation}. One can show that
\begin{equation}
\frac1{g_\rho(\tau/N)}\equiv \etabox{\mbox{\scriptsize 1}\!}{g} = \Tr_{g}(q^{L_0'-a})\ ,
\end{equation}
where $\textrm{Tr}_g$ denotes a trace (over oscillator modes in a generic Fock space) in the $g$-twisted sector with $g=1$ indicating the untwisted sector and $a$ denotes the ground state zero-point energy in the $g$-twisted sector. In the examples that we consider,
$a=1/N$, where $N$ is the order of $g$. Further, as explained in appendix \ref{derivation}, $L_0'$ is the contribution of the left-moving oscillator modes to $L_0$,

\begin{table}[hbt]
\centering
\newcommand\T{\rule{0pt}{2.6ex}}
\begin{tabular}{ccccccccc} \hline
$N$ & 1 & 2& 3 & 4 & 5 & 6 & 7 & 8 \\[2pt] \hline 
$\rho$ & $1^{24}$ &  $1^82^8$ & $1^63^6$ & $1^42^24^4$ & $1^45^4$ & $1^22^23^26^2$ & $1^37^3$ & $1^2 2^14^18^2$ \T \\[3pt] \hline
conj. class & 1A & 2A & 3A & 4B & 5A & 6A & 7A/B & 8 A \\[3pt] \hline
\end{tabular}
\caption{Cycle shapes and their $M_{24}$ conjugacy classes\cite{Atlasv3} for the $\mathbb{Z}_N$-orbifold.}\label{cycleshapes}
\end{table}

\subsection{From eta-products to dyon degeneracies}

One of the standard constructions of Siegel modular forms is through the additive (Saito-Kurokawa-Maa\ss\cite{Maass,EichlerZagier}) lift and its variants\cite{Gritsenko:1999,GritsenkoNikulinI,GritsenkoNikulinII,Gritsenko:2008}. All of them take a (weak) Jacobi form of $\Gamma_0^J(N)$ as input and give a Siegel modular form of the same weight as output. We shall indicate the additive lift symbolically by
\begin{equation}
\Phi_k(\mathbf{Z}) = \mathcal{A}\Big[\phi_{k,1}(\tau,z)\Big]\ ,
\end{equation}
where the \textit{additive seed} $\phi_{k,1}(\tau,z)$ is a Jacobi form of weight $k$, index $1$ of a suitable subgroup of the Jacobi group, possibly with character.  We will see that the Jacobi form is determined completely by the eta-products that generate the counting of $\tfrac12$-BPS states.

A physical derivation of the Siegel modular form $\Phi_k(\mathbf{Z})$ (in description 3 of Figure \ref{CHLdualitychain}) from the D1-D5 system was carried out by David-Sen\cite{David:2006yn}. The result is obtained as a product of three distinct contributions in their setup. One has
\begin{equation}
\frac{64}{\Phi_k(\mathbf{Z})} = \frac{4\ \eta(\tau)^6}{\theta_1(\tau,z)^2} \times 
\frac{16}{g_\rho(\tau/N)}\times \frac1{\mathcal{E}\!\left(K3/\mathbb{Z}_N\right)}\ ,
\end{equation}
where the first term arises from the overall motion  of the D1-D5 system in Taub-NUT space; the second term arises from the excitations of KK monopoles\footnote{In description two, the excitations of the KK monopole gets mapped to states of the heterotic/CHL string.} and third term, $\mathcal{E}\!\left(K3/\mathbb{Z}_N\right)$, arises from the excitations of the D1-branes (wrapping $S^1$) moving inside the D5-brane wrapping $K3\times S^1$) -- this is the second-quantized elliptic-genus of $K3/\mathbb{Z}_N$\cite{Dijkgraaf:1996xw}. It was observed in \cite{Govindarajan:2009qt} that the inverse of the product of the first two terms is the (weak) Jacobi form 
\begin{equation}
\widetilde{\phi}_{k,1}(\tau,z) =  \frac{\theta_1(\tau,z)^2}{\eta(\tau)^6} \times 
g_\rho(\tau/N) \ .
\end{equation}
However, $g_\rho(\tau/N)$ is not a modular form of $\Gamma_0(N)$ but is rather a modular form of the conjugate group $S \cdot \Gamma_0(N)\cdot S^{-1}$, where $S:\tau \rightarrow -1/\tau$. Hence, the S-transform of $g_\rho(\tau/N)$ which is proportional to $g_\rho(\tau)$ gives rise to a different Jacobi form\footnote{This Jacobi form will reappear in the next section when we construct modular forms  twisted BPS states in the heterotic string on $T^6$.}
\begin{equation}
\phi_{k,1}(\tau,z) =  \frac{\theta_1(\tau,z)^2}{\eta(\tau)^6} \times 
g_\rho(\tau) \ .
\end{equation}
This Jacobi form gives rise to a Siegel modular form using the additive lift\cite{Maass,Jatkar:2005bh,Govindarajan:2009qt}
\begin{align}
\Phi_k(\mathbf{Z}) &= \mathcal{A}\Big[\phi_{k,1}(\tau,z)\Big]\ , \nonumber \\
& = \sum_{\substack{n,m \in \BZ_{>0} \\ \ell \in\BZ \\ (4nm-\ell^2)>0}}\ \  \sum_{\substack{d | (n,\ell,m)\\ (d,N)=1}} \ \chi(d)\  
d^{k-1}\ a\left(\tfrac{nm}{d^2},\tfrac\ell{d}\right)\ 
q^{n} r^{\ell} s^{m}\ , \label{twistedHet}
\end{align}
where the non-trivial characters are
$$
\chi(d)=\left\{
\begin{array}{lcl}
\left(\tfrac{-1}{~d}\right), &\textrm{for} & N=4\ , \\[5pt]
\left(\tfrac{-7}{~d}\right), &\textrm{for} &N=7\ ,\\[5pt]
\left(\tfrac{-2}{~d}\right) &\textrm{for} & N=8\ .
\end{array}\right.
$$

Following \cite{Jatkar:2005bh,David:2006ud,Govindarajan:2009qt},we define $\widetilde{\Phi}_k(\mathbf{Z})$ 
 by the \textit{S-transform}
 ($\textrm{vol}_\rho \equiv  \prod_{j=1}^N (j)^{a_j}$) 
  \begin{equation}\label{Stransform1}
 \widetilde{\Phi}_k(\mathbf{Z}) \equiv (\textrm{vol}_\rho)^{1/2} \ \tau^{-k} \
\Phi_k(\mathbf{\widetilde{Z}})\ ,
\end{equation}
with
\[
\tilde{\tau} = -1/\tau \quad,\quad \tilde{z} = z/\tau\quad, \quad 
\tilde{\sigma} = \sigma -z^2/\tau\ .
\]
Explicit formulae for $\widetilde{\Phi}_k(\mathbf{Z})$ in some cases have been derived in Appendix C  of \cite{Jatkar:2005bh}.  The Fourier expansion of $\widetilde{\Phi}_k(\mathbf{Z})$ about the cusp at $\tau,  \sigma=i\infty$ is  thus equivalent to the Fourier expansion of $\Phi_k(\mathbf{Z})$ about an inequivalent cusp at $\tau=0,\  \sigma=i\infty$.\\

\noindent
\textbf{Note:} We will denote the two Siegel modular forms  $\Phi_k(\mathbf{Z})$ and $\widetilde{\Phi}_k(\mathbf{Z})$ (that were constructed in this subsection) respectively by  $\Phi^{(1,N)}_k(\mathbf{Z})$  and $\Phi^{(N,1)}_k(\mathbf{Z})$  to better organize the many Siegel modular forms that appear in the sequel.

\section{Modular forms counting twisted BPS states}

\subsection{A twist in the dyon partition function}

The inverse of the modular forms that we considered in the previous section arise from four-dimensional  indices given  by helicity traces, $B_4$ and $B_6$, which are defined as follows\cite{Gregori:1997hi}
\begin{equation}
B_{2n} = \frac1{2n!} \textrm{Tr}\left[(-1)^{2h} \ (2h)^{2n}\right]\ ,
\end{equation}
where $h$ is the third component of the angular momentum of a state in the rest frame and trace is over all states carrying a given set of charges. The  index $B_6$ receives contributions from states that break $12$ of $16$ supersymmetries while the index $B_4$ receives contributions from states that break $8$ of the supersymmetries. The inverse of the modular forms, $\Phi_k(\mathbf{Z})$ and $g_\rho(\tau/N)$, are the generating functions associated, respectively, with the indices $B_6$ and $B_4$.

Sen recently proposed a generalization of this index\cite{Sen:2009md}. Let $g$ be a symmetry of the theory -- this requires one to restrict the moduli to be on special subspaces compatible with this symmetry. Further, one restricts the charges to be $g$-invariant. The twisted indices defined by Sen are\cite{Sen:2009md} 
\begin{equation}
B^g_{2n} = \frac1{2n!} \textrm{Tr}\left[g\ (-1)^{2h} \ (2h)^{2n}\right]\ .
\end{equation}
For CHL models, we will construct modular forms that arise as generating functions for these twisted indices using the additive lift.  Sen has constructed the same modular forms using the D1-D5 system (as in \cite{David:2006yn}) thereby obtaining product formulae for these modular forms\cite{Sen:2009md,Sen:2010ts}. Thus, our results complement the results of Sen.

\subsection{Generating functions for twisted degeneracies}

Let $h$ generate a symplectic automorphism of K3 of order $M\leq4$ (under which all sixteen supersymmetries are invariant) and further it  commutes with the $Z_N$-orbifolding in the CHL model. We will refer to states which contribute to the twisted index  $\mathbb{Z}_M$-twisted states. Let us denote by $d^h(n)$ (resp.  $D^h(n,\ell,m)$) the degeneracy of $\mathbb{Z}_M$-twisted electrically charged $\tfrac12$-BPS states (resp. dyonic $\tfrac14$-BPS states). We define generating functions as we did for the untwisted case. 
\begin{align}
\frac{16}{g_\rho(\tau/N)} &= \sum_{n} d^h(n)\ q^{n/N}\ , \nonumber \\
\frac{64}{\Phi^{(N,M)}(\mathbf{Z})} &= \sum_{(n,\ell,m)} D^h(n,\ell,m)\ q^{n/N} r^\ell s^m\ .
\end{align}
We will again find that these two functions turn out to be modular forms as in the untwisted case. Further, the $\tfrac12$-BPS counting leads to eta-products associated with cycle shapes $\rho$ -- we have indicated this by adding a subscript $\rho$ to the generating function.

\subsection{Eta-products for twisted $\tfrac12$-BPS states}

We have seen that electrically charged $\tfrac12$-BPS states correspond to states of the CHL string with the supersymmetric (right-moving) sector in their ground state. The level matching condition then relates the oscillator level of left-movers to $\mathbf{q}_e^2$. Generalizing the discussion to the twisted case, we see that the following trace is the generating function of $\tfrac12$-BPS multiplets
\begin{equation}
\frac1{g_\rho(\tau/N)}\equiv \etabox{\mbox{\textit{\scriptsize h}}\!}{g} = \Tr_{g}(h\ q^{L_0'-a})\ ,
\end{equation}
where $L_0'$ is the contribution of the left-moving oscillator modes as explained in appendix \ref{derivation}.

\subsubsection*{Heterotic string on $T^6$}

For the heterotic string on $T^6$ ($N=1$ and $g=1$), the generating functions, $g_\rho(\tau)$, are given by multiplicative eta-products from the same cycle shapes given in Table \ref{cycleshapes} that appeared in the untwisted computation for the CHL $\BZ_M$-orbifolds. This is easy to see since the two computations are related by $\tau \rightarrow -1/\tau$ transformation up to a numerical factor. 

\subsubsection*{CHL $\BZ_N$-orbifolds}

For the CHL $\BZ_N$-orbifolds, a straightforward  computation shows that the result is expressible again as multiplicative eta-products arising from cycle shapes of the 24-dimensional permutation 
representation of the Mathieu group, $M_{24}$ -- see Table \ref{twistedcycleshapes} where we summarize the results. Unlike the cycle shapes that appeared in the untwisted counting, these do  \textit{not} reduce to conjugacy classes of  the Mathieu group, $M_{23}$. This is easily seen as none of the cycle shapes have length-one cycles. However, they do consist of at least five orbits.
\begin{table}[h]
 \newcommand\T{\rule{0pt}{2.6ex}}
 \centering
\begin{tabular}{ccccccc}\hline
$G$ & $H$ & $\rho$ & conj. class & $g_\rho(\tau/N)$  & $k$ &$\chi(d)$\\[2pt] \hline 
\multirow{2}{*}{$\mathbb{Z}_2$ } & $\mathbb{Z}_2$ & $2^{12}$& 2B & $\eta(\tau)^{12}$& $4$ &\T \\ \cline{2-7}
  & $\mathbb{Z}_4$ & $2^44^4$& 4A &$\eta(\tau)^{4}\eta(2\tau)^4$ & $2$&\T \\ \hline
$\mathbb{Z}_3$  & $\mathbb{Z}_3$ & $3^8$ & 3B & $\eta(\tau)^{8}$ &$2$& \T \\[3pt] \hline
\multirow{2}{*}{$\mathbb{Z}_4$ }  & $\mathbb{Z}_4$ & $4^6$ & 4C & $\eta(\tau)^{6}$ & $1$&$ \left(\tfrac{-4}{~d}\right) \T$ \\[3pt] \cline{2-7}
  & $\mathbb{Z}_2$ & $2^44^4$& 4A & $\eta(\tau)^{4}\eta(\tau/2)^4$ & $2$ & \T \\ \hline
\end{tabular}
\caption{The eta-products (along with cycle shapes, $M_{24}$ conjugacy classes\cite{Atlasv3}, weight $k$ of the Siegel modular form)  that count  $h$-twisted $\tfrac12$-BPS dyons in the CHL model for the group $G$.}\label{twistedcycleshapes}
\end{table}

\subsection{From eta-products to twisted-dyon degeneracies}

 Following the conjecture in \cite{Govindarajan:2009qt}, we consider the additive lift of the following (weak) Jacobi form with $g_\rho(\tau/N)$ taken  from Table \ref{twistedcycleshapes}.
\begin{equation}
\phi_{k,1}(\tau,z) \equiv \frac{\theta_1(\tau,z)^2}{\eta(\tau)^6} \times 
g_\rho(\tau/N) = \sum_{n,\ell} a(n,\ell)\ q^{n/N} r^\ell \ .
\end{equation}

\subsubsection*{Heterotic string on $T^6$ $(N=1)$}

The additive lift is exactly the one we considered in Eq. \eqref{twistedHet} and thus the Siegel modular forms $\Phi^{(1,M)}(\mathbf{Z})$ that we constructed are the (proposed) generating functions for $\BZ_M$-twisted dyons in the heterotic string compactified on $T^6$. This is in agreement with the results of Sen where possible and extends it in other cases -- the additive lift given here complementing the product formulae naturally appearing in Sen's approach\cite{Sen:2009md}.

\subsubsection*{CHL models with $N=M>1$}

We shall focus first, for simplicity, on the situation when $N=M=2,3,4$ as they lend a uniform construction.
The additive lift of the  Jacobi forms that appear have been considered in work of Gritensko and Nikulin\cite{GritsenkoNikulinII} where several generalizations of the Maa\ss\ additive lift have been considered. In particular, we will consider the one called Lift$_1(\phi_{k,1})$ in \cite{GritsenkoNikulinII}. This lift leads to a Siegel modular form, $\Phi_k(\mathbf{Z})$,  of weight $k$ of the double extension, $\Gamma^+_N$, of the paramodular group, $\Gamma_N$, with character induced by $(v_\eta)^{24/N}\times 1$ (see appendix \ref{character}). The formula for $\Phi_k(\mathbf{Z})$ given by the additive lift is
\begin{equation}
\Phi^{(N,N)}_k(\mathbf{Z})\equiv \sum_{\substack{n,m\in\BZ_+; \ell\in \BZ \\ n,m=1\textrm{ mod } N\\  4nm -N\ell^2 \geq 0}}\ \  
\sum_{\substack{ d  \mid (n,\ell,m)}} \ \chi(d)\  
d^{k-1}\ a\left(\tfrac{nm}{d^2},\tfrac\ell{d}\right)\ 
q^{n/N} r^{\ell} s^{m}\ ,
\end{equation}
where $a(n,\ell)$ are the Fourier coefficients of the Jacobi form and $\chi(d)$ is as given in Table \ref{twistedcycleshapes}. These three modular forms are the squares of the modular forms denoted $\Delta_{k/2}(\mathbf{Z})$ by Gritsenko and Nikulin in \cite{GritsenkoNikulinII,Gritsenko:1999a} -- the easy way to see this is to see that the squares of Jacobi forms  of index $1/2$  that appear in  the additive lift for  $\Delta_{k/2}(\mathbf{Z})$  are precisely the additive seeds for $\Phi_k(\mathbf{Z})$ (for $k=4,2,1$). We can show that $\Delta_2(\mathbf{Z})$ (and $\Phi^{(2,2)}_{4}(\mathbf{Z})$) can be written as a product of four genus-two theta constants (defined in appendix \ref{thetadefs}) as shown
\begin{equation}
\Phi^{(2,2)}_{4}(\mathbf{Z})=\left(\frac1{16}\ \theta\!\left[\begin{smallmatrix} 0\\ 1 \\ 0 \\ 0\end{smallmatrix}\right]\!\!\left(\mathbf{Z}\right)\ 
\theta\!\left[\begin{smallmatrix} 1\\ 1 \\ 0 \\ 0\end{smallmatrix}\right]\!\!\left(\mathbf{Z}\right)\ 
\theta\!\left[\begin{smallmatrix} 0\\ 1 \\ 1 \\ 0\end{smallmatrix}\right]\!\!\left(\mathbf{Z}\right)\ 
\theta\!\left[\begin{smallmatrix} 1\\ 1 \\ 1 \\ 1\end{smallmatrix}\right]\!\!\left(\mathbf{Z}\right)\ 
\right)^2\equiv \left[\Delta_2(\mathbf{Z})\right]^2\ .
\end{equation}
The $q\leftrightarrow s^2$ symmetry of $\Phi^{(2,2)}_{4}(\mathbf{Z})$ gives a second equivalent representation in terms of different theta constants:
\begin{equation}
\Phi^{(2,2)}_{4}(\mathbf{Z})=\left(\frac1{16}\ \theta\!\left[\begin{smallmatrix} 1\\ 0 \\ 0 \\ 0\end{smallmatrix}\right]\!\!\left(\mathbf{Z}'\right)\ 
\theta\!\left[\begin{smallmatrix} 1\\ 1 \\ 0 \\ 0\end{smallmatrix}\right]\!\!\left(\mathbf{Z}'\right)\ 
\theta\!\left[\begin{smallmatrix} 1\\ 0 \\ 0 \\ 1\end{smallmatrix}\right]\!\!\left(\mathbf{Z}'\right)\ 
\theta\!\left[\begin{smallmatrix} 1\\ 1 \\ 1 \\ 1\end{smallmatrix}\right]\!\!\left(\mathbf{Z}'\right)\ 
\right)^2\equiv \left[\Delta_2(\mathbf{Z})\right]^2\ .
\end{equation}
where $\mathbf{Z}'=\begin{pmatrix}  \tau/2 &  z \\ z & 2 \sigma\end{pmatrix}$.

Similarly, $\Delta_{1/2}(\mathbf{Z})$ can be written in terms of a single genus-two theta constant as follows:
\begin{equation}
\Phi^{(4,4)}_{1}(\mathbf{Z})= \left(\frac12 \theta\!\left[\begin{smallmatrix} 1\\ 1 \\ 1 \\ 1\end{smallmatrix}\right]\!\!\left(\mathbf{Z}'\right)\ 
\right)^2\equiv\left[\Delta_{1/2}(\mathbf{Z})\right]^2\ ,
\end{equation}
where $\mathbf{Z}'=\begin{pmatrix}  \tau & 2 z \\ 2z & 4 \sigma\end{pmatrix}$.  

\subsubsection*{CHL models with $N\neq M$}

The remaining two modular forms, $\Phi^{(2,4)}_2(\mathbf{Z})$ and $\Phi^{(4,2)}_2(\mathbf{Z})$,  need a generalization of the additive lift for paramodular groups to higher level. This has been considered recently by Clery and Gritsenko\cite{Gritsenko:2008}. These are modular forms of the doubly extended paramodular group, $\Gamma^+_2(2)$ at level two. The relevant cycle shape given in Table \ref{twistedcycleshapes} is $2^44^4$ and the corresponding Jacobi form is
\begin{equation}
\phi_{2,1}(\tau,z)= \frac{\theta_1(\tau,z)^2}{\eta(\tau)^6} \times \eta(\tau)^4 \eta(2\tau)^4 = \sum_{\substack{n\in \BZ_{>0}, \ell\in \BZ\\ n=1\textrm{ mod }2 \\[1pt] 2n - \ell^2 > 0}} a(n,\ell)\ q^{n/2} r^\ell \ .
\end{equation}
The additive lift of the above Jacobi  form is given by
\begin{align}
\Phi^{(2,4)}_2(\mathbf{Z})&=\mathcal{A}\Big[\phi_{2,1}(\tau,z)\Big] \nonumber \\
&\equiv \sum_{\substack{n,m\in\BZ_+; \ell\in \BZ \\ n,m=1\textrm{ mod } 2\\  2nm -\ell^2 > 0}}\ \  
\sum_{\substack{ d  \mid (n,\ell,m)\\ (d,2)=1}} \ 
d\ a\left(\tfrac{nm}{d^2},\tfrac\ell{d}\right)\ 
q^{n/2} r^{\ell} s^{m}\ .
\end{align}
Again, one can show that $\Phi^{(2,4)}_2(\mathbf{Z})$ is the square of the modular form, $Q_1$, defined in \cite{Gritsenko:2008}. This more or less follows from the relationship between the Jacobi forms that generate them as well as their modular properties.

The modular form $\Phi^{(4,2)}_2(\mathbf{Z})$ corresponds to $\Phi^{(2,4)}_2(\mathbf{Z})$  expanded about the other cusp. It is defined through the S-transform as follows 
\begin{equation}\label{Stransform2}
\Phi^{(4,2)}_2(\mathbf{Z}) \equiv 4 \ \tau^{-2} \
\Phi^{(2,4)}_2(\mathbf{\widetilde{Z}})\ ,
\end{equation}
with
\[
\tilde{\tau} = -1/\tau \quad,\quad \tilde{z} = z/\tau\quad, \quad 
\tilde{\sigma} = \sigma -z^2/\tau\ .
\]
We will denote the square-root of by $\Phi^{(4,2)}_2(\mathbf{Z})$  by $\widetilde{Q}_1$ -- this can equivalently be defined in terms of $Q_1$ expanded about the other cusp and hence is a well-defined Siegel modular form. We have shown that both $Q_1$ and $\widetilde{Q}_1$ can be defined in terms of genus-two theta constants (defined in appendix \ref{thetadefs}). We obtain the following expressions
\begin{align}
\Phi^{(2,4)}_2(\mathbf{Z})  &= \left(\frac14\ \theta\!\left[\begin{smallmatrix} 1\\ 0 \\ 0 \\ 1\end{smallmatrix}\right]\!\!\left(\mathbf{Z}'\right)\ 
\theta\!\left[\begin{smallmatrix} 1\\ 1 \\ 1 \\ 1\end{smallmatrix}\right]\!\!\left(\mathbf{Z}'\right)\ 
\right)^2=\left[Q_1(\mathbf{Z})\right]^2\ ,\\
\Phi^{(4,2)}_2(\mathbf{Z})  &= \left(\frac12\ \theta\!\left[\begin{smallmatrix} 0\\ 0 \\ 1 \\ 1\end{smallmatrix}\right]\!\!\left(\mathbf{Z}'\right)\ 
\theta\!\left[\begin{smallmatrix} 1\\ 1 \\ 1 \\ 1\end{smallmatrix}\right]\!\!\left(\mathbf{Z}'\right)\  \right)^2=\left[\widetilde{Q}_1(\mathbf{Z})\right]^2\ ,
\end{align}
where $\mathbf{Z}'=\left(\begin{matrix} \tau & 2z \\ 2z & 4 \sigma \end{matrix}\right)$. The $q\leftrightarrow s^2$ symmetry of $\Phi^{(2,4)}_2(\mathbf{Z})$ gives a second representation in terms of genus-two theta constants.
\begin{equation}
\Phi^{(2,4)}_2(\mathbf{Z})  = \left(\frac14\ \theta\!\left[\begin{smallmatrix} 0\\ 1 \\ 1 \\ 0\end{smallmatrix}\right]\!\!\left(2\mathbf{Z}\right)\ 
\theta\!\left[\begin{smallmatrix} 1\\ 1 \\ 1 \\ 1\end{smallmatrix}\right]\!\!\left(2\mathbf{Z}\right)\ 
\right)^2=\left[Q_1(\mathbf{Z})\right]^2\ .
\end{equation}
We provide a partial Fourier expansion for $Q_1$ and $\widetilde{Q}_1$ in appendix \ref{explicit}.

\subsection{Properties of the modular forms}

In this section we have constructed Siegel modular forms, $\Phi^{(N,M)}(\mathbf{Z})$, that generate the degeneracy of $\BZ_M$-twisted dyons in the CHL $\BZ_N$-orbifold for all values of $N,M\leq 4$. We now list some of the properties of $\Phi^{(N,M)}(\mathbf{Z})$ with $N\leq M$ (when $N>M$, the modular forms are given by the $S$-transform given in Eqs. \eqref{Stransform1} and \eqref{Stransform2})
\begin{enumerate}
\item They are modular forms of the paramodular group $\Gamma_t(P)$ with level $P=M/(N,M)$ and $t=(N,M)$. 
\item Clery and Gritsenko define dd-modular forms to be all Siegel modular forms that vanish exactly on the discriminant-one Humbert surface $\mathcal{H}_1\equiv ( \mathbf{Z}\in \mathbb{H}_2| z=0)$ and its $\Gamma_t(P)$ translates with order one and  classified them\cite{Gritsenko:2008}. All modular forms that we have constructed are squares of dd-modular forms.  We will summarise the detailed relationship in the Table~\ref{periodictable}. 
\item  As all our modular forms are squares of  dd-modular forms,  they have double zeros at $z=0$ and its images under the action of $\Gamma_t(P)$. One has
$$
\lim_{z\rightarrow 0}\Phi^{(N,M)}(\mathbf{Z}) = (2\pi z)^2 g_\rho(\tau/N)\  g_\rho(\sigma)\ ,
$$
where $\rho$ is the relevant cycle shape.
\item The extended S-duality group $\widehat{\Gamma}_1(N)$ ($N=tP'$)  can be embedded in $\Gamma_t^{(2)}(P)$ as follows:
\begin{equation}\label{Sdualityembedding}
  \Gamma_0^{(1)}(tP')\ni \gamma\equiv \begin{pmatrix} a & b \\ c t & d \end{pmatrix} \longmapsto 
\widehat{\gamma} \equiv \begin{pmatrix} d & -c t & c & 0 \\-b & a & 0 & b/t \\0 & 0 & a & b \\ 0 &
0 & c t & d\end{pmatrix} \in \Gamma_t(P)\ ,
\end{equation}
with $c=0 \textrm{ mod } P'$. This generalizes the embeddings considered earlier\cite{Jatkar:2005bh,Govindarajan:2008vi} to the paramodular group. The modular forms are invariant under extended S-duality i.e.,
\begin{equation}
\Phi^{(N,M)}(\widehat{\gamma}\cdot \mathbf{Z}) = \Phi^{(N,M)}(\mathbf{Z})\ .
\end{equation}
This is a consequence of the character, $v(\widehat{\gamma})=1$, for all the generators of the extended S-duality group -- the details are provided in the appendix \ref{Sdualityproof}.
\item  The formula for the additive lift(s) clearly shows that the modular forms have a symmetry corresponding to $q \leftrightarrow s^N$ (or $\tau \leftrightarrow N \sigma$) -- this corresponds to exchanging the definition of electric and magnetic charges. This can be stated as the the invariance of the modular form under the action of the generator $V_t$ defined in Eq. \eqref{Vtdef}.
\begin{equation}
\Phi^{(N,M)}(V_t\cdot \mathbf{Z}) = \Phi^{(N,M)}(\mathbf{Z})\ .
\end{equation}
\item The degeneracies of the $\BZ_M$-twisted quarter BPS states in the CHL $\BZ_N$-orbifolds  are given by an integral\cite{Jatkar:2005bh}
\begin{equation}\label{degeneracyintegral}
D^h(n,\ell,m) = \frac{(-1)^\ell}{N} \int_C d\mathbf{Z}\  \frac{e^{-2\pi i\Tr(\mathcal{Q}\cdot \mathbf{Z})}}{\Phi^{(N,M)}(\mathbf{Z})}\ ,
\end{equation}
where $\Tr(\mathcal{Q}\cdot \mathbf{Z}) = n\tau/N + \ell z + m \sigma$ and $C$ is a three-dimensional subspace given by (we write $\tau=\tau_1+i\tau_2$, $z=z_1+i z_2$ and $\sigma=\sigma_1+i\sigma_2$)
$$
0\leq \tau_1 \leq N \quad, \quad 0\leq z_1 \leq 1 \quad,\quad 0\leq \sigma_1 \leq 1 \ ,
$$
with their imaginary parts being fixed to  large positive numbers.
\item For large charges, i.e., $\tfrac{1}2\mathbf{q}_e^2\gg 0$, $\mathbf{q}_m^2\gg 0$ and $\mathbf{q}_e\cdot \mathbf{q}_m\gg 0$, Sen shows that the macroscopic entropy is given by\cite{Sen:2009md,Sen:2010ts} 
\begin{equation}
S_{\textrm{BH}}= \frac1M ~\Big[\pi \sqrt{\mathbf{q}_e^2~\mathbf{q}_m^2-(\mathbf{q}_e\cdot \mathbf{q}_m)^2}\Big]\stackrel{?}{=}S^h_{\textrm{stat}}  \equiv  \log D^h.
\end{equation}
This is smaller than the entropy for untwisted states by a factor of $M$. One can ask whether the statistical entropy given matches the above answer. It is clear that the leading contribution arises from the dominant saddle-point in the integral \eqref{degeneracyintegral} -- this corresponds to identifying the dominant zero of the modular form. For the modular forms $\Phi^{(M,M)}(\mathbf{Z})$, according to \cite[Eq. 1.22]{Gritsenko:2008} the zeros occur at the quadratic rational divisor
\begin{equation}
t\ n_2(\tau \sigma-z)^2 + t n_1 \sigma + j z +  m_1\tau +m_2=0\ , (j,m_1,m_2,n_1,n_2)\in \BZ\ ,
\end{equation}
when the discriminant  $D=j^2 -4t m_1n_1-4t m_2n_2 =1$. Using the computation described in \cite{Dijkgraaf:1996it,Cardoso:2004xf,Jatkar:2005bh,Sen:2007qy}, one obtains that the dominant zero occurs when $|n_2|=1$ and the answer matches the macroscopic computation, as $t=(M,M)=M$. Sen has also verified this relation in several examples, in particular  $\Phi^{1,M}(\mathbf{Z})$ for which $t=1$ and $P=M$\cite{Sen:2009md,Sen:2010ts}. It appears that the zeros of all modular forms are of the form\footnote{The only two examples for which this has not been proved are $(N,M)=(2,4),(4,2)$.}
\begin{equation}
t P\ n_2(\tau \sigma-z)^2 + (\textrm{ terms linear in } \tau, z, \sigma) =0\ ,\quad n_2 \in \BZ\ ,
\end{equation}
with discriminant one. The dominant contribution is always from $|n_2|=1$ and gives a result that is consistent with the macroscopic answer on using $M=tP$.
\item All the modular forms admit product formulae with even multiplicities. This is important for the BKM superalgebra interpretation that we discuss next. The detailed formulae are discussed in  appendix \ref{productformula}. 
\end{enumerate}
The properties of the modular forms are thus consistent with macroscopic considerations discussed in \cite{Sen:2009md, Sen:2010ts}. This provides substantial evidence that the conjecture for the additive seed given in \cite{Govindarajan:2009qt}.

\section{BKM Lie superalgebras}

The main idea  in \cite{Govindarajan:2008vi} and subsequently in \cite{Cheng:2008kt, Govindarajan:2009qt} was to associate an algebraic structure to  the square-root of the Siegel modular forms that  generate the degeneracy of  $\tfrac14$-BPS CHL dyons. This was carried out in these papers by  showing that  the square-root of these generating functions are related  the Weyl-Kac-Borcherds (WKB) denominator formula of a BKM Lie superalgebra. The intent of this section is to extend these considerations to the modular forms that were considered in the previous section.  In other words, we will look for a family of BKM Lie superalgebras, that we denote by $\mathcal{G}_N(M)$, whose WKB denominator formula is given by 
$$\Delta_{k/2}^{(N,M)}(\mathbf{Z})\equiv \sqrt{\Phi^{(N,M)}_k(\mathbf{Z})}\ .$$
Since we do not provide an introduction to BKM Lie superalgebras, we refer the reader to \cite{Ray:2006,Nikulin:1995} for a mathematical introduction to BKM Lie superalgebras as well \cite{Govindarajan:2008vi,Cheng:2008kt} for a physical introduction in the context of counting dyons. All the BKM Lie superalgebras that appear in this paper are of a special kind -- they are all (Borcherds)  extensions of rank-three Lorentzian Kac-Moody Lie superalgebras with only even(bosonic) simple real roots. These Lie algebras have been studied extensively and classifed by Gritensko and Nikulin(see \cite{Gritsenko:2002} and references therein).

\subsection{Wall crossing: a prelude to BKM Lie superalgebras}

Following a very nice observation due to Cheng and Verlinde in \cite{Cheng:2008fc}, we expect  that the walls of marginal stability of the $\tfrac14$-BPS dyons get mapped to the walls of the Weyl chamber of the BKM Lie superalgebra $\mathcal{G}_N(M)$. For the BKM Lie algebras $\mathcal{G}_N(1)$ with $N\leq 4$ that are associated with untwisted dyons, this has been been verified in \cite{Cheng:2008kt,Govindarajan:2009qt}.

Sen argues that since the twisted dyons occur in the same CHL model, the S-duality group as well as the walls of marginal stability should remain unchanged for the twisted dyons. Assuming that the BKM Lie superalgebra $\mathcal{G}_N(M)$ for $M>1$ exists, then the physical prediction is that  its Weyl chamber is identical to that of the the BKM Lie superalgebra $\mathcal{G}_N(1)$. Since every wall in the Weyl chamber is identified with a simple real root, it implies that, for fixed $N$ and varying $M$ the Lie superalgebras $\mathcal{G}_N(M)$ should have the \textit{same} set of real simple roots and Cartan matrix. 

In \cite{Govindarajan:2008vi}, the existence of the the BKM Lie superalgebras $\mathcal{G}_1(M)$ for $M=1,2,3,4,5$ was established. Indeed, it was shown that all the five BKM Lie superalgebras shared the same Weyl vector, three simple real roots and Cartan matrix $A^{(1)}$ defined in Eq. \eqref{Cartanmatrices}. One sees that this agrees with the physical expectations. Does it hold when $M>1$?

\subsection{Studying the structure of $\mathcal{G}_N(M)$}

First, $\Delta_{k/2}^{(N,M)}(\mathbf{Z})$ must be a well-defined modular form. As we have seen in this previous section, that all of them turn out to be dd-modular forms in the classification of Clery and Gritsenko\cite{Gritsenko:2008}. So this is trivially satisfied in all our examples and we list them in Table \ref{periodictable}. The dd-modular forms should satisfy the following properties for them to lead to a BKM Lie superalgebra -- we will restrict our remarks to the  dd-modular forms, $\Delta_2$, $\Delta_1$, $\Delta_{1/2}$, $Q_1$ and $\widetilde{Q}_1$, in this section as all other examples have already been considered in the physics literature(see \cite{Govindarajan:2008vi,Cheng:2008kt,Govindarajan:2009qt}.)
\begin{itemize}
\item We should be able to provide representations in terms of sums and products for them. The additive lifts provide the sum side of the denominator identity. Product formulae for four of the dd-modular forms in the form of multiplicative (Borcherds) lifts of weight zero modular forms of index $(N,M)$ have already appeared in the literature\cite{GritsenkoNikulinII,Gritsenko:2008} and they are given in appendix \ref{productformula}.
\item We need to establish the integrality of all coefficients. On the sum side, it follows from our observation that $\Delta_2$, $\Delta_{1/2}$, $Q_1$ and $\widetilde{Q}_1$ have representations as products of genus-two theta constants (the dividing factors of two cancel out). On the product side, we also need to show that the multiplicities of various roots are integers. This follows from the fact that the weak Jacobi forms that are used in the multiplicative lift used in \ref{productformula} all have integral coefficients\cite{GritsenkoNikulinII,Gritsenko:2008}.
\item We need to show that the simple real roots all appear with correct multiplicity. This is achieved by showing the invariance of the modular forms under $\gamma^{(N)}$ and $\delta$ as they generate all the simple real roots through their action. 
Since we have already proven the invariance of $ \Phi_{k}^{(N,M)}( \mathbf{Z})$ under S-duality, we need to verify that no sign arises on taking the square-root. A practical way to do this is to check to see that there is no relative sign between two terms that are related by the action of  $\gamma^{(N)}$ and $\delta$. We find
\begin{equation}
\Delta_{k/2}^{(N,M)}(\gamma^{(N)}\cdot \mathbf{Z})= \Delta_{k/2}^{(N,M)}( \mathbf{Z})\quad \textrm{and}\quad \Delta_{k/2}^{(N,M)}(\delta \cdot \mathbf{Z})= \Delta_{k/2}^{(N,M)}( \mathbf{Z})\ .
\end{equation}
\item The sum side of the denominator formula implies that $\Delta_{k/2}^{(N,M)}( \mathbf{Z})$ must change sign under a Weyl reflection due to any simple real root. Due to the dihedral  symmetry  of the  modular form, it suffices to show that the modular form is an odd function of $z$ :
$$
\Delta_{k/2}^{(N,M)}(w\cdot\mathbf{Z})\equiv \Delta_{k/2}^{(N,M)}\big(\left(\begin{smallmatrix} \tau & -z \\ -z & \sigma\end{smallmatrix}\right)\big) = - \Delta_{k/2}^{(N,M)}\big(\left(\begin{smallmatrix} \tau & z \\ z & \sigma\end{smallmatrix}\right)\big)
$$
The $z\rightarrow -z$ operation is an elementary Weyl reflection due to a simple real root present in all the models. This implies that the extended S-duality group is realized in the Lie algebra as
\begin{equation}
\widehat{\Gamma}_1(N) = \mathcal{W}(A^{(N)})\rtimes \textrm{Dih}(\mathcal{P}_N)\ ,
\end{equation}
where $ \mathcal{W}(A^{(N)})$ is the Weyl group generated by all the simple real roots and Dih$(\mathcal{P}_N)$ is the dihedral group acting on the fundamental polygon, $\mathcal{P}_N$  that represents the Weyl chamber\cite{Cheng:2008fc,Cheng:2008kt,Govindarajan:2009qt}.
\end{itemize}

\subsection{The BKM Lie superalgebras: $\mathcal{G}_N(M)$}

The considerations of the previous subsection leads to the  existence of a family of BKM Lie superalgebras, $\mathcal{G}_N(M)$, ($N=1,2,3,4$)  satisfying the following properties:
\begin{enumerate}
\item The BKM Lie superalgebras $\mathcal{G}_N(M)$ arise as inequivalent (for different values of $M$) automorphic extensions of the rank-three Lie algebra with Cartan matrix $A^{(N)}$. In other words, for a given  $N$, all the BKM Lie superalgebras $\mathcal{G}_N(M)$ have identical real simple roots as well as Weyl vector. However, the imaginary simple roots differ.
\begin{align}\label{Cartanmatrices}
A^{(1)} &=A_{1,II}=\begin{pmatrix}
~~2 & -2 & -2 \\
-2 & ~~2 & -2 \\
-2 & -2 & ~~2
\end{pmatrix}\quad,\quad 
\\
A^{(2)} &=A_{2,II} \equiv \begin{pmatrix} ~~2 & -2 &-6 &-2\\[-3pt] -2 & ~~2&-2& -6\\[-3pt]-6& -2&~~2&-2\\[-3pt] -2& -6& -2&~~2 \end{pmatrix} \ ,
\\
A^{(3)} &=A_{3,II} \equiv \begin{pmatrix} ~~2 & -2 &-10 &-14&-10 &-2\\[-3pt] -2 & ~~2&-2& -10& -14&-10\\[-3pt] -10& -2&~~2&-2&-10&-14\\[-3pt] -14& -10& -2&~~2&-2&-10\\[-3pt] -10 &-14&-10 &-2& ~~2 &-2\\[-3pt] -2 &-10 &-14&-10 &-2&~~2\end{pmatrix} \ ,
\\[4pt]
A^{(4)} &= (a_{nm})\quad \textrm{where}\quad 
a_{nm}= 2 -4(n-m)^2\  , \textrm{ with } m,n\in \mathbb{Z}\ .
\end{align}
In the above equations, $A_{N,II}$, is the notation used by Gritsenko and Nikulin in their classification of Cartan matrices for  rank-three Lorentzian Kac-Moody Lie algebras.
\item The Weyl-Kac-Borcherds denominator formula for $\mathcal{G}_N(M)$ correspond to the product and sum representations of a Siegel modular form $\Delta^{(N,M)}_{k/2}(\mathbf{Z})$ -- the weight $k$ is dependent on both $N$ and $M$. 
\item  The square of the Siegel modular form,  $\Delta_{k/2}^{(N,M)}(\mathbf{Z})$, is the generating function of $\mathbb{Z}_M$-twisted dyons in the  CHL $\BZ_N$-orbifold. The allowed values of $M$ correspond to  symplectic involutions of $K3$ of the form $\BZ_M\times \BZ_N$.
\item The walls of the Weyl chamber of $\mathcal{G}_N(M)$ are independent of $M$ and get mapped to the  walls of marginal stability of the dyons.
\end{enumerate}
It is easy to check that all the aforementioned properties follow from the properties of the modular forms $\Phi_k^{(N,M)}(\mathbf{Z})$ discussed in the previous section up to signs that can be fixed by considering terms appearing in the Fourier expansions of the dd-modular forms $\Delta_{k/2}^{(N,M)}(\mathbf{Z})$. A more direct approach is to prove the properties directly from the modular properties of the dd-modular form. The table \ref{periodictable} provides a nice summary of the results and constitutes the main result of this paper. 

\textbf{Remarks:}  The BKM Lie superalgebras $\mathcal{G}_N(1)$ and $\mathcal{G}_1(M)$ were denoted respectively by
$\widetilde{\mathcal{G}}_N$ and $\mathcal{G}_M$ in \cite{Govindarajan:2008vi,Govindarajan:2009qt}. The BKM Lie superalgebras $\mathcal{G}_N(N)$ and their Cartan matrix were discussed in \cite{GritsenkoNikulinII}. Thus the BKM Lie superalgebras associated with $Q_1$ and $\widetilde{Q}_1$ are new though their occurence was anticipated in \cite{Gritsenko:2008}.

\begin{table}
 \newcommand\T{\rule{0pt}{3.2ex}}
 \centering
\begin{tabular}{c||c|c|c|c||c}\hline
\backslashbox{$N$}{$M$} & $1$ & $2$ & $3$ & $4$ & Cartan matrix \\ \hline \hline
$1$ & $\stackrel{1^{24}}{\Delta_5}$ & $\stackrel{1^82^8}{\nabla_3}$ & $\stackrel{1^63^6}{\nabla_2}$ & $\stackrel{1^42^24^4}{\nabla_{3/2}}$\T & $A^{(1)}$ \\[3pt] \hline
$2$ & $\stackrel{1^82^8}{\widetilde{\nabla}_3}$ &$\stackrel{2^{12}}{\Delta_2}$ & \xcancel{\phantom{\T$\nabla_{3/2}$}} & $\stackrel{2^{4}4^4}{Q_1}$\T & $A^{(2)}$ \\[3pt] \hline
$3$ & $\stackrel{1^63^6}{\widetilde{\nabla}_2}$ &\xcancel{\phantom{\T$\nabla_{3/2}$}}& $\stackrel{3^{8}}{\Delta_1}$\T & \xcancel{\phantom{\T$\nabla_{3/2}$}} & $A^{(3)}$\\[3pt] \hline
$4$ & ~$\stackrel{1^42^24^4}{\widetilde{\nabla}_{3/2}}$ & $\stackrel{2^{4}4^4}{\widetilde{Q}_1}$&\xcancel{\phantom{\T$\nabla_{3/2}$}} & $\stackrel{4^{6}~}{\Delta_{1/2}}$\T& $A^{(4)}$ \\[3pt] \hline
\end{tabular}
\caption{The periodic table of  BKM Lie superalgebras, $\mathcal{G}_N(M)$: the the $(N,M)$ entry is the relevant cycle shape and the  dd-modular form in the notation of Clery-Gritsenko \cite{Gritsenko:2008} -- the dd form is  $\Delta_k^{N,M}(\mathbf{Z})$ in our notation. Note that all BKM Lie superalgebras in a given row have identical  Cartan matrices as specified in the last column. The modular forms $\nabla_k$ and $\widetilde{\nabla}_k$ are related by the S-transform lead to distinct BKM Lie superalgebras.}
\label{periodictable}
\end{table}
\section{Concluding Remarks}

We have seen that the counting of twisted BPS states in CHL models has lead to a nice connection with dd-modular forms as well as rank-three Lorentzian Kac-Moody algebras. The BKM Lie superalgebras $\mathcal{G}_4(M)$ for $M=1,2,4$ provide us concrete examples of rank-three Lorentzian Kac-Moody Lie algebras of parabolic type -- they all have an infinite number of simple real roots that physical considerations imply \textit{must} be identical. A practical result of our study is that several of these dd-modular forms can be written in terms of products of genus-two theta constants. It appears that one can write product formulae for all the ten even genus-two theta constants starting from the most odd even theta-constant.\footnote{It appears that this has been known to Gritsenko for several years and must not be considered as an original observation!}

It is hard to miss the fact that the Mathieu group, $M_{24}$, played an important role in the construction of the Siegel modular forms. Multiplicative eta products identical to the ones that appeared in the counting of $\tfrac12$-BPS states played an important role in constructing twisted versions of the Fake Monster Lie algebra of Borcherds\cite{BorcherdsFakeMonster,ScheithauerConway1,Niemann}. It is natural to ask if the Lie algebras $\mathcal{G}_N(M)$ are twisted versions of the Lie algebra $\mathcal{G}_N(1)$ and it appears to be the case. It is also of interest to ask whether modules of the BKM Lie superalgebras $\mathcal{G}_N(1)$ decompose into irreps of $M_{24}$  leading to a \textit{moonshine} for $M_{24}$ analogous to the one for the Monster group. An exciting paper by Cheng\cite{Cheng:2010pq} (that appeared recently) explores this aspect as well the appearance of the $M_{24}$ in the elliptic genus of $K3$ (and hence on the product side of these modular forms) in the work of Eguchi et al.\cite{Eguchi:2010ej}(see also \cite{Gaberdiel:2010ch}). In work that is appear soon, using replication formulae, we show that the Siegel modular forms $\Phi^{(1,M)}(\mathbf{Z})$ (and their inverses) are obtained as twisted traces of  a $M_{24}$-module\cite{Govindarajan:2011}.

In their classification of rank-three Lorentzian Kac-Moody Lie algebras\cite{Gritsenko:2002}, Gritsenko and Nikulin observe that there are three kinds of such algebras based on the behavior of the fundamental chamber, $\mathcal{P}$, of the Weyl group and the Weyl vector, $\rho$. Those of \textit{elliptic type} have  $\rho^2<0$ and $\mathcal{P}$ has finite volume; those of \textit{parabolic type} have  $\rho^2=0$  while those of \textit{hyperbolic type} have $\rho^2>0$. In the other two cases, the Weyl chamber has finite volume only under some restrictions (we refer the reader to \cite{Gritsenko:2002} for further details). 
We restricted our considerations to CHL $\mathbb{Z}_N$-orbifolds with $N\leq 4$ in this paper. The models with $N=1,2,3$ leads to algebras of elliptic type while the $N=4$ models lead to algebras of parabolic type. 
However, it appears that, at the very least, the $N=5$ model might lead to the first example of a rank-three Lorentzian Kac-Moody Lie algebra of hyperbolic type as the Weyl vector has $\rho^2>0$ and the candidate Weyl chamber as obtained from the walls of marginal stability appear to of the correct type\cite{Krishna:2010gc}. The associated modular form is expected to be meromorphic and thus has not appeared in the list of dd-modular forms of  Clery-Gritsenko\cite{Gritsenko:2008}. It is of interest to study other kinds of twisted dyons in CHL models -- it is possible to consider twists that do \textit{not} commute with the $\BZ_N$-orbifold  such as $\BZ_2$ actions that lead to dihedral groups that are symplectic automorphisms of K3. Another possibility, is to look for twists that break supersymmetry -- these might give insights into BPS state counting in models with $\mathcal{N}=2$ supersymmetry. \\[6pt]

\noindent 
\textbf{Acknowledgments:} We  thank Prof. V.~Gritsenko, D.~Jatkar and K.~Gopala Krishna for useful conversations.  We also thank the organizers of the National Strings Meeting 2010(NSM10) held at IIT Bombay (Feb. 10-15, 2010) as well as the Workshop on  ``Automorphic forms, Kac-Moody Lie algebras and Strings" held Max Planck Institut f\"ur Mathematik, Bonn (May 10-14, 2010) for the opportunity to present these results at the workshops.  This manuscript was completed during a visit to the Albert Einstein Institut, Golm and we thank Stefan Theisen and all members of AEI for a wonderful atmosphere.

\appendix

\section{Cycle shapes to eta-products: a derivation}\label{derivation}

This appendix provides a derivation and more importantly, the intuition behind the appearance of multiplicative eta products in counting $\tfrac12$-BPS states. In particular, we establish the map, Eq. \eqref{cycleshapemap}, that directly relates cycle shapes to multiplicative eta products.

\subsubsection*{Counting states of the CHL string}

Electrically charged $\tfrac12$-BPS states arise as states of the heterotic string. For the CHL-$\mathbb{Z}_N$ orbifolds, electrically charged states carrying fractional charge such that $\tfrac{N}2 \mathbf{q}_e^2 \in \mathbb{Z}$ arise in the $g$-twisted sector, where $g^N=1$. Thus, we will carry out the counting in the $g$-twisted sector i.e.,
$$
\mathbf{X}(\sigma+2\pi) = g\cdot \mathbf{X}(\sigma)\ ,
$$
where $\mathbf{X}$ represents the $24$ left-moving scalars in the bosonic sector of the heterotic string in the light-cone gauge and $\sigma$ is the circle coordinate on the worldsheet with cylindrical topology. Recall that $g$ has no action on the supersymmetric right-movers as well as the two remaining left-movers arising from four-dimensional spacetime.

In the type IIA picture (Description 1), the action of $g$ involves a shift on the $S^1$ combined with a symplectic automorphism of $K3$. The shift affects the zero-modes (momenta and winding modes along the circle) but does not affect on the oscillator modes. Using the duality that relates  the heterotic string to the NS5-brane wrapped on $K3$, the action on the scalars can be worked out. In fact, it suffices to know the cycle shapes corresponding to the action of $g$ on $H^*(K3,\mathbb{Z})$ since $g$ acts trivially on the right-movers. The action on the oscillator (non-zero) modes have the \textit{same} cycle shape. 

Following the arguments of Sen\cite[see section 3]{Sen:2005ch}, one can show that the level-matching condition for a $\tfrac12$-BPS state with charge $\tfrac{1}2 \mathbf{q}_e^2$ is given by
\begin{equation}
 \tfrac{1}2 \mathbf{q}_e^2 = L_0' +a \ ,
\end{equation}
where $a$ is the contribution of zero-point energies from the chiral bosons and $L_0'$ represents the contributions of the left-moving oscillator modes\footnote{In an arbitrary twisted sector, $L_0'$ also includes the contribution of momenta that don't contribute to the electric charge (as the charge lattice may be only a sub-lattice in   momentum lattice in the twisted sector) -- this is not the case in the twisted sector that we consider. For instance, in the untwisted sector of a CHL model, the allowed  momenta lie in the $(22,6)$ dimensional lattice of the heterotic string on $T^6$ though the CHL models have charges valued in some sub-lattice.}. 
Thus, we need to count the degeneracy of $\mathbb{Z}_N$-invariant states of oscillator number $\tfrac{1}2 \mathbf{q}_e^2 + a$ in a $g$-twisted sector. The generating function of these degeneracies are given by
\begin{equation}
\frac1N \ \sum_{s=0}^{N-1} \textrm{Tr}_g \left(g^s\ q^{L_0'+a}\right)\ . 
\end{equation}
where the trace runs over all excitations of left-moving oscillator modes and the sum imposes the projection on to $\mathbb{Z}_N$-invariant states. This expression can be simplified since the $\mathbb{Z}_N$-projection implements level matching in the twisted sector.  Since level-matching has already been imposed, $g$ acts trivially on these states -- we an replace the sum by a single term given by the $s=0$ piece and removing the factor of $\tfrac1N$. We finally obtain the following result: the degeneracy of the electric $\tfrac12$-BPS is generated by 
\begin{equation}
\etabox{\mbox{\scriptsize 1}\!}{g} \equiv \textrm{Tr}_g \left(\ q^{L_0'+a}\right)\ . 
\end{equation}

\subsubsection*{Computing the zero-point energy}

 A boson with  fractional moding $\alpha_{n+\phi}$ (with $0\leq \phi < 1$) gives rise to  a zero-point energy contribution given by $a= \big[ - \tfrac1{16} (2\phi-1)^2+\tfrac1{48}\big]$. 
 For twisted bosons that are part of a cycle of length $m$,  the moding (in a diagonal basis) is given by $\phi_j=j/m\textrm{ mod }1$ for $j=0,1,\ldots, m-1$. The cumulative contribution of the $m$-cycle to the zero-point energy is $\tfrac{-1}{24m}$. The zero-point energy for $24$ bosons in cycle shapes given in table \ref{cycleshapes} for the CHL $\BZ_N$-orbifold is then equal to $a=-1/N$.

\subsubsection*{Deriving the eta-products}

Let $\phi_j$ denote the eigenvalues of $g$ acting on the $24$ light-cone scalars. The contribution of the oscillators from these scalars to $\Tr_g~q^{L_0'}$ is
\begin{equation}
\etabox{\mbox{\scriptsize 1}\!}{g} = q^{a} \Big[ \prod_{j=1}^{24} \prod_{n=0}^\infty \!{}^{'}\left(1- q^{n+\phi_j}\right)\Big]^{-1} \ ,
\end{equation}
where the prime over the product indicates that the $n=0$ term is excluded for all $j$ for which  $\phi_j=0$. 
In order to understand the appearance of eta products, consider $m$ scalars that form a cycle of length $m$ under the action of $g$. The $m$ eigenvalues of $g$ are thus $\phi_j = j/m$ for $j=0,1,\ldots, (m-1)$ which gives rise to the product that contributes to $\Tr_g~q^{L_0'}$:
\begin{equation}
\Big[ \prod_{j=1}^{m} \prod_{n=0}^\infty \left(1- q^{n+j/m }\right)\Big]^{-1} = \Big[ \prod_{n=1}^\infty \left(1- q^{n/m}\right)\Big]^{-1}\propto \Big[\eta\left(\tfrac{\tau}m\right)\Big].
\end{equation}
It is easy to verify that the factor of $q^{a}$ provides precisely the power of $q$ needed to convert the  Euler functions to an eta function.

For counting twisted BPS states, we need obtain the simultaneous eigenvalues under the action of $g$ and $h$ -- call them $\phi_j$ and $\psi_j$ respectively. Then, the only change in the earlier computation that lead to eta products is the addition of phases $\exp(2\pi i\psi_j)$ leads to the following formula:
\begin{equation}
\etabox{\mbox{\textit{\scriptsize h}}\!}{g} = q^{a} \Big[ \prod_{j=1}^{24} \prod_{n=0}^\infty\!{}^{'} \left(1- e^{2\pi i \psi_j}\ q^{n+\phi_j}\right)\Big]^{-1} \ ,
\end{equation}
where the prime over the product indicates that the $n=0$ term is excluded for all $j$ for which  $\phi_j=0$. 
We illustrate the computation for a specific example when  both $g$ and $h$ generate $\BZ_2$ -- both of them have cycle shapes $1^82^8$ since the cycle shape is uniquely determined in terms of the order of the element.  We now need to specify explcitly the action of $g$ and $h$ -- this needs more details given, for instance, by Chaudhuri and Lowe. We quote their result where they write out the action of $g$ and $h$ on the scalars. $g$ acts with eigenvalues $(1^8, (-1)^8,1^8)$ and $h$ acts with eigenvalues $(1^8,(-1)^4,1^4, (-1)^4, 1^4)$. Again, we track only terms that contribute to $\Tr_g~h~q^{L_0'}$ and obtain
\begin{align}
 &q^{-1/2}\ \Big[\prod_{n=0}^\infty\!{}^{'} \left(1- q^{n}\right)^8 \left(1+ q^{n+\tfrac12}\right)^4 \left(1- q^{n+\tfrac12}\right)^4 \left(1+ q^{n}\right)^4 \left(1-  q^{n}\right)^4\Big]^{-1}  \nonumber \\
 &=q^{-1/2}\ \Big[\prod_{n=0}^\infty\!{}^{'} \left(1- q^{n}\right)^8 \left(1- q^{2n+1}\right)^4 \left(1-  q^{2n}\right)^4\Big]^{-1} \\
 &=\ \Big[q^{1/24}\prod_{n=1}^\infty \left(1- q^{n}\right)^{12}\Big]^{-1} = \eta(\tau)^{-12} = \Big[g_{\rho}(\tfrac{\tau}2)\Big]^{-1}\textrm{ for } \rho=2^{12}\ . \nonumber
\end{align}
A straightforward and mildly tedious computation (not shown here) leads to the remaining eta products quoted in Table \ref{twistedcycleshapes}. We can carry a couple of consistency checks on results. First, the leading power of $q$ coming from the eta products always reduce to $1/N$ as required from considerations of zero-point energies.
Second, for the $\BZ_N\times \BZ_N$ examples,  the eta products take the form $\eta(\tau)^D$ (for some $D$) -- this is related to the fact that $\etabox{\mbox{\textit{\scriptsize h}}\!}{g}=\etabox{\mbox{\textit{\scriptsize g}}\!}{h}$ in these instances. 

\section{The paramodular group}\label{modularstuff}

The group $Sp(2,\mathbb{Q})$ is the set of $4\times 4$ matrices written 
in terms of four $2\times 2$ matrices $A,\, B,\, C,\, D$ (taking values in $\mathbb{Q}$)
as
$$
M=\begin{pmatrix}
   A   & B   \\
    C  &  D
\end{pmatrix}
$$
satisfying $ A B^T = B A^T $, $ CD^T=D C^T $ and $ AD^T-BC^T=I $. 
This group acts naturally 
on the Siegel upper half space, $\BH_2$, as
\begin{equation}
\mathbf{Z}=\begin{pmatrix} \tau & z \\ z & \sigma \end{pmatrix}
\longmapsto M\cdot \mathbf{Z}\equiv (A \mathbf{Z} + B) 
(C\mathbf{Z} + D)^{-1} \ .
\end{equation}

The paramodular group at level $P$ that we denote by $\Gamma_t(P)$ is defined as follows (we follow \cite{Gritsenko:2008} for all definitions) (for $t,P\in \mathbb{Z}_{>0}$):
\begin{equation}
\Gamma_t(P) = \left\{
\left(\begin{smallmatrix}   
* & *t & * & * \\[2pt] * & * & * & *t^{-1} \\[2pt] *P & *Pt & * & * \\[2pt] *Pt & *Pt & *t & *  
\end{smallmatrix}\right)\in \textrm{Sp}(2,\mathbb{Q}),\ \textrm{all } * \in \BZ \right\}\ .
\end{equation}
When $t=1$, then $\Gamma_1=\Sp(2,\BZ)\equiv \Gamma^{(2)}$ is the usual  symplectic group and $\Gamma_1(P)=\Gamma^{(2)}_0(P)$ is its congruence subgroup at level $P$. Further when $P=1$, we get the full paramodular group, $\Gamma_t$.

We denote by $\Gamma^+_t(N)=\Gamma_t(N)\cup \Gamma_t(N) V_t$  a normal double extension of $\Gamma_t(N)$ in $\Sp(2,\mathbb{R})$ with\footnote{We will denote by $\Gamma^+_t(1)$ by $\Gamma_t^+$.}
\begin{equation}\label{Vtdef}
V_t = \tfrac1{\sqrt{t}} \left(\begin{smallmatrix} 0 & t & 0 & 0 \\ 1 & 0 & 0& 0 \\
0 & 0 & 0 & 1 \\ 0 & 0 & t & 0 \end{smallmatrix}\right)\ ,
\end{equation}
with $\det(CZ+D)=-1$.
This acts on $\BH_2$ as 
\begin{equation}
(\tau,z,\sigma ) \longrightarrow (t \sigma, z, \tau/t)\ .
\end{equation}
The group $\Gamma^+_t(P)$ is generated by $V_t$ and its parabolic subgroup
\begin{equation}
\Gamma_t^\infty(P) = \left\{
\left(\begin{smallmatrix}   
* & 0 & * & * \\[2pt] * & 1 & * & *t^{-1} \\[2pt] *P & 0& * & * \\[2pt] 0 & 0 & 0 & 1  
\end{smallmatrix}\right)\in \Gamma_t(P),\ \textrm{all } * \in \BZ \right\}\ .
\end{equation}
The Jacobi group  is defined by 
\begin{equation}\label{JacobiGroup}
\Gamma^J(P)=\big(\Gamma_t^\infty(P) \cap \Sp(2,\BZ)\big)/{\pm \mathbf{1}_4} \simeq \Gamma_0^{(1)}(P)\ltimes H(\BZ)\ .
\end{equation} 
The embedding of 
$\left(\begin{smallmatrix} a & b \\ c & d\end{smallmatrix}\right)
\in \Gamma_0^{(1)}(P)$ in $\Gamma_t(P)$ is given by
\begin{equation}
\label{sl2embed}
\widetilde{\begin{pmatrix} a & b \\ c & d \end{pmatrix}}
\equiv \begin{pmatrix}
   a   &  0 & b & 0   \\
     0 & 1 & 0 & 0 \\
     c &  0 & d & 0 \\
     0 & 0 & 0 & 1  
\end{pmatrix}
\ , \ c=0 \mod P \ .
\end{equation}
The above matrix acts on $\BH_2$ as
\begin{equation}
(\tau,z,\sigma) \longrightarrow \left(\frac{a \tau + b}{c\tau+d},\  
\frac{z}{c\tau+d},\  \sigma-\frac{c z^2}{c \tau+d}\right)\ ,
\end{equation}
with $\det(C\mathbf{Z} + D)=(c\tau +d)$. The Heisenberg group, 
$H(\BZ)$, is generated by $Sp(2,\BZ)$ matrices of the form
\begin{equation}
\label{sl2embedapp}
[\lambda, \mu,\kappa]\equiv \begin{pmatrix}
   1   &  0 & 0 & \mu   \\
    \lambda & 1 & \mu & \kappa \\
     0 &  0 & 1 & -\lambda \\
     0 & 0 & 0 & 1  
\end{pmatrix}
\qquad \textrm{with } \lambda, \mu, \kappa \in \BZ
\end{equation}
The above matrix acts on $\BH_2$ as
\begin{equation}
(\tau,z,\sigma) \longrightarrow \left(\tau,\ z+ \lambda \tau  + \mu,\  
\sigma + \lambda^2 \tau + 2 \lambda z + \lambda \mu +\kappa \right)\ ,
\end{equation}
with $\det(C\mathbf{Z} + D)=1$. It is easy to see that $\Gamma^J$ 
preserves the one-dimensional cusp at $\textrm{Im}(\sigma)= \infty$.

\subsection{Characters of $\Gamma_t^+(P)$}

A \textit{Siegel modular form} of weight $k$ and character $v$ 
with respect to $\Gamma_t(P)$ is a holomorphic function $F: \BH_2 
\rightarrow \BC$ satisfying
\begin{equation}
F(M\cdot \mathbf{Z}) = v(M)\ \det(C\mathbf{Z}+D)^k \ F(\mathbf{Z})\ ,
\end{equation} 
for all $\mathbb{Z}\in \BH_2$ and $M\in \Gamma_t(P)$.  Let $M_1$ and $M_2$ be any two  $\Gamma_t(P)$ matrices. Then, for any character one has
\begin{equation}
v(M_1\cdot M_2) = v(M_1)\  v(M_2)\ .
\end{equation}
This property is useful in simplifying the computation of characters into those that generate the group.

We wish to derive the characters of the modular forms constructed by the additive lift in section 3.
The character is induced by the character of the Jacobi form that is the additive seed. Hence we first discuss the characters of the Jacobi forms before establishing the characters for the Siegel modular forms.

\subsubsection*{Characters of the additive seeds}\label{character}

The Jacobi forms (of index $t$) are modular forms of the Jacobi group $\Gamma^J(P)\simeq \Gamma_0^{(1)}(P)\ltimes H(\BZ)$. It suffices to give the character, $\chi$, for $\Gamma^{(1)}_0(P)$ transformations and character $v_H^{2t}$ under the Heisenberg group -- we indicate this by $\chi \times v_H^{2t}$. Recall that 
\begin{equation}
v_H\big([\lambda,\mu,\kappa]\big) = (-1)^{\lambda+\mu+\lambda\mu+\kappa}\ ,
\end{equation}
is the unique binary character of the Heisenberg group. A useful observation for our purposes is that weight $-1$, index 1 Jacobi form $\big[\vartheta_1(\tau,z)/\eta(\tau)^3\big]$ has character $1\times v_H$. The additive seed for modular forms $\Phi_k^{(N,M)}(\mathbf{Z})$ is given by
\begin{equation}
\phi_{k,1}(\tau,z) = \left[\frac{\vartheta_1(\tau,z)^2}{\eta(\tau)^6}\right]\times g_\rho(\tau/N)\ . 
\end{equation}
The character of the additive seed is thus \textit{completely} determined by the eta product $g_\rho(\tau/N)$ since the first factor transforms without character as $v_H^2=1$. When $N=M=1,2,3,4$, since $g_\rho = \eta(\tau)^{24/N}$, the character is given by $\chi_\rho\equiv v_\eta^N$ where $v_\eta$ is the character of the Dedekind eta function which is always a $24$-th root of unity -- an explicit formula may be found, for instance, in \cite{Dummit:1985,GritsenkoNikulinII}.. 
Using the details of $v_\eta$, one can show that 
\begin{equation}
\chi_\rho(T) = e^{2\pi i/N}\textrm{ and } \chi(\gamma)=1 \textrm{ for } \gamma= \begin{pmatrix} 1 & 0 \\ N & 1\end{pmatrix}\ .
\end{equation}
For $N=2$, $M=4$ one has\cite{Gritsenko:2008}
\begin{equation}
\chi_\rho(\gamma)=(-1)^ {d(b-c)}\ \textrm{ for }\ \gamma=\begin{pmatrix} a & b \\ 2 c & d\end{pmatrix} \in \Gamma^{(1)}_0(2)\ . 
\end{equation}

\subsubsection*{Characters of the Siegel modular form}

The results quoted below are taken from the additive lifts  for paramodular groups\cite[Theorem 1.12]{GritsenkoNikulinII} and their congruence subgroups\cite[Theorem 2.2]{Gritsenko:2008}.  Let $\phi$ be a Jacobi form  with weight $k$, index $1$  whose additive lift (Lift${}_1$) gives a Siegel modular form, $\Phi$ of the paramodular group $\Gamma_t(P)$. Since $\Gamma_t^+(P)$ is generated by $V_t$ and $\Gamma_t^\infty(P)$,  the characters for the $V_t$ and $\Gamma_t^\infty(P)$ completely specify the character of $\Phi$ under arbitrary elements of $\Gamma_t^\infty(P)$. One has 
\begin{equation}
v(V_t)= (-1)^k\ ,
\end{equation} 
where $k$ is the weight of the modular form.
This can also be seen by noticing the $q\leftrightarrow s^N$ symmetry of the modular forms. $\Gamma_t^\infty(P)$ differs from $\Gamma^J(P)$ by its center -- these are elements of the form $[0,0,\kappa/t]$. Thus, the character of elements in $\Gamma_t^\infty(P)$ of the form 
 $[\chi\times 1 \times e^{2\pi i \kappa /t}]$
 where $[\chi\times 1]$ is the character of the Jacobi form $\phi$ and $e^{2\pi i \kappa/t}$ is the character of the element $[0,0,\kappa/t]$.

\noindent \textbf{Conclusion:} The character of the Siegel modular form, $\Phi^{(N,M)}(\mathbf{Z})$, is of the  form $[\chi_\rho\times 1 \times e^{2\pi i \kappa /t}]$,
where $\chi_\rho$ is the character induced by the eta product $g_\rho(\tau/N)$ with $t=(N,M)$. In other words, consider an element of $\Gamma_t(P)$ of the form $U=\widetilde{\gamma}\cdot [\lambda,\mu,\kappa']\cdot[0,0,\kappa/t]$, then $v(U)=\chi(\gamma)\times e^{2\pi i \kappa /t}$ where $\widetilde{\gamma}$ is related to $\gamma$ through Eq. \eqref{sl2embed}.

\subsection{S-duality invariance of $\Phi_k^{(N,M)}(\mathbf{Z})$}\label{Sdualityproof}

We need to show that the modular forms $\Phi_k^{(N,M)}(\mathbf{Z})$ under the three generators, $\gamma^{(N)}$, $\delta$ and $w$ embedded into $\Gamma_t(P)$ as described in Eq. \eqref{Sdualityembedding}. 
\begin{itemize}
\item The action of $w$ on $\mathbf{Z}$ is given by $z\rightarrow -z$ with $\tau$ and $\sigma$ being invariant. The only $z$ dependence in the additive seed appears through $\vartheta_1(\tau,z)^2$ which is an even function of $z$. It is easy to see from the detailed formula for the additive lift that it implies that all  $\Phi_k^{(N,M)}(\mathbf{Z})$ are even functions of $z$. A similar argument shows that their square-roots are necessarily odd functions of $z$. 
\item Next, consider the generatory $\delta$ or equivalently, $w\cdot \delta$. The generator  $w\cdot \delta$ gets mapped to the element $[-1,0,0]$ of $H(\BZ)$ and hence the modular forms are all invariant as the additive seeds transform without character under $H(\BZ)$.  
\item For $(N=2, M=4)$, one can show that $\chi_\rho(\gamma^{(2)})=1$.  For the others instances with $N=M$,  first observe that one can write 
$$
\gamma^{(N)} = \begin{pmatrix} 1 & 0 \\ N & 1 \end{pmatrix}\cdot T^{-1}\quad  \textrm{ where }\ T= \begin{pmatrix} 1 & 1 \\ 0 & 1 \end{pmatrix} \ .
$$
Thus one has 
$$
\chi_\rho\left(\gamma^{(N)}\right) = \chi_\rho\
\Big(\left(\begin{smallmatrix} 1 & 0 \\ N & 1 \end{smallmatrix}\right)\Big) \times \chi_\rho(T^{-1})=\chi_\rho(T^{-1})=\frac1{\chi_\rho(T)}\ .
$$ 
It suffices to study the character induced by $T$ in the Siegel modular form.
One can show that the embedding of $T$ in the paramodular group, denoted by $\widehat{T}$ (Eq. \eqref{Sdualityembedding}),  is given by the following product 
\begin{equation}
\widehat{T}=V_t\cdot U_0 \cdot \widetilde{T^{-1}}\cdot V_t\cdot \widetilde{T}\cdot V_t\cdot U_0^{-1}\cdot V_t\ ,
\end{equation}
where $\widetilde{T}$ is defined in Eq. \eqref{sl2embed} and
$$
U_0 = \begin{pmatrix} 1 & 0 & 0 & 0 \\
 0 & 0 & 0 & \tfrac{1}{t} \\
 0 & 0 & 1 & 1 \\
 t & -t & 0 & 0
 \end{pmatrix}\ .
$$
The character, $v(\widehat{T})$, thus reduces to the one induced by $\chi_\rho(T^{-1})\times \chi_\rho(T)=1$. An alternate method to fix $v(\widehat{\gamma^{(N)}})$ is to check the relative sign (or possible phase) of any two elements such as two real simple roots that are related by $\gamma^{(N)}$ action -- we have used this method for $\widetilde{Q}_1$, for instance.
\end{itemize}
This completes that proof of the invariance of the modular forms under extended S-duality.






\section{Product formulae for $\Phi^{(N,M)}_k(\mathbf{Z})$}

\subsection{Theta Functions}\label{thetadefs}

The genus-one theta functions are defined by
\begin{equation}
\theta\left[\genfrac{}{}{0pt}{}{a}{b}\right] \left(\tau,z\right)
=\sum_{l \in \BZ} 
q^{\frac12 (l+\frac{a}2)^2}\ 
r^{(l+\frac{a}2)}\ e^{i\pi lb}\ ,
\end{equation}
where $a.b\in (0,1)\mod 2$. We define $\vartheta_1 
\left(\tau,z\right)\equiv\theta\left[\genfrac{}{}{0pt}{}{1}{1}\right](\tau,z)$,
$\vartheta_2 
\left(\tau,z\right)\equiv\theta\left[\genfrac{}{}{0pt}{}{1}{0}\right] 
\left(z_1,z\right)$, $\vartheta_3 
\left(\tau,z\right)\equiv\theta\left[\genfrac{}{}{0pt}{}{0}{0}\right] 
\left(\tau,z\right)$ and $\vartheta_4 
\left(\tau,z\right)\equiv\theta\left[\genfrac{}{}{0pt}{}{0}{1}\right] 
\left(\tau,z\right)$.

We define the genus-two theta constants as follows\cite{Nikulin:1995}:
\begin{equation}
\theta\left[\genfrac{}{}{0pt}{}{\mathbf{a}}{\mathbf{b}}\right]
\left(\mathbf{Z}\right)
=\sum_{(l_1, l_2)\in \BZ^2} 
q^{\frac12 (l_1+\frac{a_1}2)^2}\ 
r^{(l_1+\frac{a_1}2)(l_2+\frac{a_2}2)}\ 
s^{\frac12 (l_2+\frac{a_2}2)^2}\ 
e^{i\pi(l_1b_1+l_2b_2)}\ ,
\end{equation}
where $\mathbf{a}=\begin{pmatrix}a_1\\ a_2
\end{pmatrix}$,
$\mathbf{b}=\begin{pmatrix}b_1\\ b_2
\end{pmatrix}$,
and $\mathbf{Z}=\begin{pmatrix}\tau & z \\ z &
\sigma \end{pmatrix}\in \mathbb{H}_2$.

\subsection{Weak Jacobi forms}

We will consider the following weak Jacobi modular forms of weight zero and index $t$, $\phi_{0,t}$ in constructing product formulae for the Siegel modular forms\cite[see Lemma 2.5 and Example 2.3]{GritsenkoNikulinII}. 
\begin{align}
\phi_{0,2}(\tau,z) &= \frac{\phi_{2,2}(\tau,z)}{\eta(\tau)^4} = \sum_{n,\ell\in \BZ} c(n,\ell)\ q^{n} r^{\ell}\ , \nonumber \\
&=(r+4+r^{-1}) + q(r^3-8r^2-r+16 
-r^{-1}-8 r^{-2}+r^{-3})+\mathcal{O}(q^2)\nonumber \\
\phi_{0,3}(\tau,z) &=\left(\frac{\vartheta_1(\tau,2z)}{\vartheta_1(\tau,z)}\right)^2 = \sum_{n,\ell\in \BZ} c(n,\ell) q^{n} r^{\ell} \label{weakcoeff1}\\
&=(r+2+r^{-1}) + q(-4r^3-r^2+2r+4 
+2r^{-1}-4 r^{-2}-4r^{-3})+\mathcal{O}(q^2)\nonumber \\
\phi_{0,4}(\tau,z) &= \frac{\vartheta_1(\tau,3z)}{\vartheta_1(\tau,z)}= \sum_{n,\ell\in \BZ} c(n,\ell)\ q^{n} r^{\ell}\nonumber \\
&=(r+1+r^{-1}) - q(r^4+r^3-r+2 -r^{-1}+r^{-3}+r^{-4})+\mathcal{O}(q^2) \nonumber 
\end{align}
where\footnote{Let $(\phi_1,\phi_2)$ be two Jacobi forms of weights $(k_1,k_2)$ and index $(m_1,m_2)$ respectively. Then the operation, $[\phi_1,\phi_2]\equiv (m_2\phi_1'\phi_2 -m_1 \phi_2'\phi_1)/(2\pi i)$ (with $\phi'(\tau,z)=\partial_z \phi(\tau,z)$) produces a new Jacobi form of weight $(k_1+k_2+1)$ and index $(m_1+m_2)$\cite{EichlerZagier}.}
$$\phi_{2,2}(\tau,z)=2\left[\vartheta_1(\tau,z),\frac{\vartheta_1(\tau,2z)}{\vartheta_1(\tau,z)}\right]\ .$$

The product formula for $Q_1(\mathbf{Z})$ is given  by the multiplicative lift of the following weak Jacobi form\cite{Gritsenko:2008} 
\begin{equation}
\psi_{0,2}(\tau,z) = 2 \frac{\vartheta_2(\tau,2z)}{\vartheta_2(\tau,0)} \ .
\end{equation}
Since this is a level 2 Jacobi form, we need its Fourier coefficients about the cusps at $i\infty$ and $0$. The expansions are\cite{Gritsenko:2008}
\begin{align}
\psi_{0,2}(\tau,z)&= \sum_{n,\ell\in \BZ} c_1(n,\ell) \ q^{n} r^{\ell}=(r+r^{-1}) - q(r^3-r 
-r^{-1}+r^{-3})+\mathcal{O}(q^2) \label{weakcoeff2}\\
\psi_{0,2}|_S &=\sum_{n,\ell\in \BZ} c_2(n,\ell) \ q^{n} r^{\ell} = 2 -2 q^{1/2} (r^{2}-2 +r^{-2}) - 4q(r^{2}-2 +r^{-2})\nonumber \\
&\hspace{1.4in} -8q^{3/2}(r^{2}-2 +r^{-2})+\mathcal{O}(q^2)\nonumber
\end{align}
The weak Jacobi forms have integral coefficients according to \cite{GritsenkoNikulinII,Gritsenko:2008}.
\subsection{Explicit Product formulae}\label{productformula}

We provide product formulae for the modular forms $\Delta_k$ using the Borcherds (multiplicative) lift of the weak Jacobi forms, $\phi_{0,t}(\tau,z)$ of weight zero, index $t$. Let $c(n,\ell)$ be the Fourier-Jacobi coefficients of the Jacobi form as defined in  Eq. \eqref{weakcoeff1}. For $(k,t)\in \big[(2,2),(1,3),(1/2,4)\big]$, then one has the following product representation for $\Delta_k(\mathbf{Z})$\cite{GritsenkoNikulinII} 
\begin{equation}
\Delta_k(\mathbf{Z}) = q^Ar^B s^C \prod_{(n,\ell,m)>0} \big(1-q^nr^\ell s^{tm}\big)^{c(nm,\ell)}\ ,
\end{equation}
 with
 $$
 A=\tfrac1{24}\sum_{\ell\in \BZ} c(0,\ell)\ ,\  
B=\tfrac12\sum_{\ell\in \BZ,\ell>0} \ell\ c(0,\ell) \textrm{ and } C=\tfrac14\sum_{\ell\in \BZ} \ell^2\ c(0,\ell) \ .
$$
Using the explicit values of the Fourier-Jacobi coefficients given in Eq. \eqref{weakcoeff1}, one obtains that $A=\tfrac1{2t}$ and $B=C=\tfrac12$.

The product formula for $Q_1$ needs us to consider $\psi_{0,2}(\tau,z)$ which is a Jacobi form at level two. It has two cusps and let us denote the Fourier-Jacobi coefficients at the two cusps, $i\infty$ and $0$, respectively by $c_1(n,\ell)$ and $c_2(n,\ell)$. Using the fact that the width, $h$, of the cusps are $1,2$ respectively, the formula given in \cite[Theorem 3.1]{Gritsenko:2008} reduces to
\begin{equation}
Q_1(\mathbf{Z}) = q^Ar^B s^C \prod_{(n,\ell,m)>0} \big(1-q^nr^\ell s^{2m}\big)^{c_1(nm,\ell)}\times \big(1-(q^nr^\ell s^{2m})^2\big)^{c_2(nm,\ell)}\ ,
\end{equation}
 with
 $$
 A=\tfrac1{24}\sum_{j=1}^2\sum_{\ell\in \BZ} h_j\  c_j(0,\ell)\ ,\  
B=\tfrac12 \sum_{j=1}^2\sum_{\ell\in \BZ,\ell>0} \ell\ h_j \ c(0,\ell) \textrm{ and } C=\tfrac14 \sum_{j=1}^2\sum_{\ell\in \BZ} \ell^2\ h_j \ c(0,\ell)\ .
$$ 
Using the explicit values of the Fourier-Jacobi coefficients given in Eq. \eqref{weakcoeff2}, one obtains that $A=\tfrac1{4}$ and $B=C=\tfrac12$.

\section{Explicit formulae for $Q_1$ and $\widetilde{Q}_1$}\label{explicit}

We indicate the terms corresponding to simple real roots in bold face. Note that all four of them ($r^{-1}$, $qr$, $rs^2$ and $q r^3s^2$)  appear with the coefficient $-1$ indicating a multiplicity of one. 
\begin{multline*}
Q_1=q^{1/4} r^{1/2} s^{1/2}\Big[ \Big(\left(-\mathbf{\tfrac{1}{r}}+1\right)+s^2 \left(\tfrac{1}{r^2}-\tfrac{1}{r}+1-\mathbf{r}\right)+s^4
   \left(\tfrac{1}{r^2}-r\right)+\cdots \Big) \\
   +q
   \Big(\left(\tfrac{1}{r^2}-\tfrac{1}{r}+1-\mathbf{r}\right)+\left(\tfrac{1}{r^4}-\tfrac{2}{r^3}-\tfrac{1}{r}+1+2 r^2-\mathbf{r^3}\right)
   s^2+\left(-\tfrac{1}{r^5}+\tfrac{1}{r^2}-r+r^4\right) s^4+\cdots \Big) + \cdots \Big]
\end{multline*}
As for $Q_1$, we indicate in bold face four of the simple real roots  ($r^{-1}$, $qr$, $rs^4$ and $q^3 r^7s^4$) that appear in the Fourier expansion of $\widetilde{Q}$ to the order given below.
\begin{multline*}
\widetilde{Q}_1=q^{1/8} r^{1/2} s^{1/2}\Big[\left(\left(-\mathbf{\tfrac{1}{r}}+1\right)+s^2 \left(\tfrac{2}{r}-2\right)+s^4
   \left(\tfrac{1}{r^2}-\mathbf{r}\right)+\cdots \right)+ \\
   \sqrt{q}
   \left(\left(\tfrac{2}{r}-2\right)+\left(-\tfrac{2}{r^3}+\tfrac{2}{r^2}-2 r+2 r^2\right) s^2+\left(-\tfrac{2}{r^2}+2
   r\right) s^4+\cdots \right) \\
   +q \left(\left(\tfrac{1}{r^2}-\mathbf{r}\right)+\left(-\tfrac{2}{r^2}+2
   r\right) s^2+\left(-\tfrac{1}{r^5}+r^4\right) s^4+\cdots \right) \\
   +q^{3/2} \left(\left(-\tfrac{2}{r^2}+2
   r\right)+\left(\tfrac{2}{r^4}-\tfrac{2}{r}+2-2 r^3\right) s^2+\left(\tfrac{2}{r^5}-2 r^4\right)
   s^4+\cdots \right) \\
   +q^2 \left(\left(-\tfrac{2}{r}+2\right)+\left(\tfrac{2}{r^5}-\tfrac{2}{r^4}+2 r^3-2
   r^4\right) s^2+\left(\tfrac{2}{r^2}-2 r\right) s^4\cdots \right) \\
   +q^3
   \left(\left(-\tfrac{1}{r^3}+\tfrac{2}{r^2}-2 r+r^2\right)+\left(-\tfrac{2}{r^6}+\tfrac{4}{r^3}-4 r^2+2 r^5\right)
   s^2+\left(\tfrac{1}{r^8}-\tfrac{2}{r^5}+2 r^4-\mathbf{r^7}\right) s^4+\cdots\right)+\cdots   \Big]
   \end{multline*}

\bibliography{master}
\end{document}